%% file: cmpkt.tex
\documentstyle[twoside,fleqn,epsf,advrheol]{article}

\input{defpkt.tex}

\title{\bf THE KINETIC THEORY OF DILUTE SOLUTIONS OF FLEXIBLE POLYMERS: 
HYDRODYNAMIC INTERACTION}

\author{{\bf J. Ravi Prakash}\address{Department of Chemical Engineering,
Indian Institute of Technology, \\
Madras, India, 600 036 }}

\begin{document}

\maketitle

\section{INTRODUCTION}
The rheological properties of dilute polymer solutions are commonly used in
industry for characterising the dissolved polymer in terms of its molecular 
weight, its mean molecular size, its chain architecture, the relaxation 
time spectrum, the translational diffusion coefficient and so on. 
There is therefore considerable effort world wide on developing 
molecular theories that 
relate the microscopic structure of the polymer and its interactions 
with the solvent to the observed macroscopic behavior. In this chapter, 
recent theoretical progress that has been made in the 
development of a coherent 
conceptual framework for modelling the rheological properties 
of dilute polymer solutions is reviewed. 

A polymer solute molecule dissolved in a dilute Newtonian solvent is typically
represented in molecular theories by a coarse-grained mechanical model,
while the relatively rapidly varying motions of solvent molecules 
surrounding the
polymer molecule are replaced by a random force field acting on the
mechanical model. The replacement of the complex polymer molecule with a
coarse-grained mechanical model is justified by the belief that such
models capture those large scale properties of the polymer molecule, such as
its stretching and orientation by the solvent flow field, that are
considered to be responsible for the solution's macroscopic behavior.
An example of a coarse-grained model frequently used to represent a 
flexible polymer molecule is the {\it bead-spring chain}, which is a 
linear chain of identical beads connected by elastic springs.

Progress in the development of molecular theories for dilute polymer 
solutions has
essentially involved the succesive introduction, at the molecular level,
of various physical phenomena that are considered to be responsible for the
macroscopic properties of the polymer solution. For instance, the simplest theory
based on a bead-spring model assumes that the solvent influences the motion of
the beads by exerting a drag force and a Brownian force.
Since this theory fails to predict a large number of the observed
features of polymer solutions, more advanced theories have been developed
which incorporate additional microscopic phenomena. Thus, theories have
been developed which (i) include the phenomenon of `hydrodynamic interaction'
between the beads, (ii) try to account for the finite extensibility of the polymer
molecule, (iii) attempt to ensure that two parts of the polymer chain do not
occupy the same place at the same time, (iv) consider the internal
friction experienced when two parts
of a polymer chain close to each other in space move apart,
and so on. The aim of this chapter is to present the unified framework 
within which these microscopic phenomena may be treated, and to focus in 
particular on recent advances in the treatment of the effect of 
hydrodynamic interaction. To a large extent, the notation that is used 
here is the same as
that in the treatise {\it Dynamics of Polymeric Liquids} by Bird and
co-authors~\cite{birdb}. 

\section{TRANSPORT PROPERTIES OF DILUTE SOLUTIONS} 
\label{tpds}
\subsection{Dilute solutions}
A solution is considered dilute if the polymer chains are isolated from
each other and have negligible interactions with each other. In this regime of
concentration the polymer solution's properties are determined by the nature of
the interaction between the segments of a single polymer chain with each other,
and by the nature of the interaction between the segments and the 
surrounding solvent molecules. As the concentration of polymers is increased,
a new threshold is reached where the polymer molecules begin to interpenetrate
and interact with each other. This threshold is reached at a surprisingly
low concentration, and heralds the inception of the semi-dilute regime,
where the polymer solution's properties have been found to be
significantly different. Beyond the semi-dilute regime lie concentrated
solutions and melts. In this chapter we are concerned exclusively with
the behavior of dilute solutions.

A discussion of the threshold concentration at which the semi-dilute
regime in initiated is helpful in introducing several concepts that are
used frequently in the description of polymer solutions.

A polymer molecule surrounded by solvent molecules undergoes thermal
motion. A measure of the average size of the polymer molecule is the
root mean square distance between the two ends of the polymer chain,
typically denoted by $R$. This size
is routinely measured with the help of scattering experiments, and is
found to increase with the molecular weight of the polymer chain with a
scaling law, $R \sim M^\nu$, where $M$ is the molecular weight, and
$\nu$ is the scaling exponent which depends on the nature of the
polymer--solvent interaction. In {\it good \/} solvents, solute--solvent
interactions are favoured relative to solute--solute interactions. As a
consequence the polymer chain {\it swells \/} and its size is found to scale with
an exponent $\nu = 3/5$. On the other hand, in {\it poor \/} solvents,
the situation is one in which solute--solute interactions are preferred.
There exists a particular temperature, called the {\it theta \/}
temperature, at which the scaling exponent $\nu$ changes dramatically
from 3/5 to 1/2. At this temperature, the urge to expand caused by
two parts of the chain being unable to occupy the same location
(leading to the presence of an {\it excluded volume}), is just balanced by
the repulsion of the solute molecules by the solvent molecules.

Polymer chains in a solution can be imagined to begin to interact with
each other when the solution volume is filled with closely packed
spheres representing the average size of the molecule. This occurs when
$n_p \, R^3 \approx  1$, where $n_p $ in the number of chains per unit volume. 
Since $n_p =\rho_p \, N_{\rm A} / M $, where $\rho_p$ is the polymer mass density
and $N_{\rm A}$ is Avagadro's number, it follows that polymer density at
overlap, $\rho_p^*$, scales with molecular weight as,
$\rho_p^* \sim M^{1-3\nu}.$ Polymer molecules typically have 
molecular weights
between $10^4$ and $10^6$ gm/mol. As a result, it is clear that 
the polymer solution can be considered dilute only at very low 
polymer densities. Since experimental measurements are difficult at such
low concentrations, the usual practice is to extrapolate results of
experiments carried out at decreasing concentrations to the limit of
zero concentration. For instance, in the case of dilute polymer
solutions it is conventional to report
the {\it intrinsic} viscosity, which is defined by,
\begin{equation}
\lbrack \eta \rbrack = \lim_{\rho_p \to 0} \,
{\eta_p \over \rho_p \, \eta_s}
\label{invis}
\end{equation}
where $\eta_p$ is the polymer contribution to the solution viscosity, 
and $\eta_s$ is the solvent viscosity.

\subsection{Homogeneous flows}
Complex flow situations typically encountered in polymer
processing frequently involve a combination of shearing and
extensional deformations. The response of the polymer solution to these
two modes of deformation is very different. Consequently, predicting the
rheological properties of the solution under both shear and extensional
deformation is considered to be very important in order to properly
characterise the solutions behavior. 
Rather than considering flows where both these modes of deformation are
simultaneously present, it is common in polymer kinetic theory to
analyse simpler flow situations called {\it homogeneous} flows, 
where they may be treated separately.

A flow is called homogeneous, if the rate of strain tensor,
${\dot \bgam} = (\nabla \bv) (t) + (\nabla \bv)^\dagger (t)$,
where $\bv$ is the solution velocity field, is
independent of position. In other words, the solution velocity field
$\bv$ in homogeneous flows,
can always be represented as $\bv = \bv_0 + \bk (t) \cdot {\br} $,
where $\bv_0$ is a constant vector, $\bk(t) = \nabla \bv (t) $
is a traceless tensor for incompressible fluids, and ${\br}$ is the
position vector with respect to a laboratory fixed frame of reference.
While there is no spatial variation in the rate of strain tensor 
in homogeneous flows,
there is no restriction with regard to its variation in time. Therefore, the
response of dilute solutions to {\it transient} shear and extensional flows is
also used to probe its character as an alternative means
of characterisation independent of the steady state material functions.

Two homogeneous flows, steady simple shear flow and 
small amplitude oscillatory shear flow, that are frequently used 
to validate the predictions of molecular theories which incorporate 
hydrodynamic interaction, are described briefly below. A comprehensive 
discussion of material functions in various flow situations can be 
found in the book by Bird~{\it et~al.} \cite{birda}. 

\subsection{Simple shear flows}
The rheological properties of a
dilute polymer solution can be obtained once
the stress tensor, $\btau,$ is known. The stress tensor is considered to
be given by the sum of two contributions, $\btau=\btau^s + \btau^p$, 
where $\btau^s$ is the contribution from the solvent, and
$\btau^p$ is the polymer contribution. Since the solvent is assumed to
be Newtonian, the solvent stress (using a compressive definition
for the stress tensor \cite{birda}) is given by,
$\btau^s= - \, \eta_s \, {\dot \bgam}$. The nature of the
polymer contribution $\btau^p$ in simple shear flows is
discussed below.

Simple shear flows are described by a velocity field,
\begin{equation}
v_x={\dot \gamma_{yx}} \, y, v_y= 0, v_z=0
\label{sfvel}
\end{equation}
where the velocity gradient ${\dot \gamma_{yx}}$ can be a function of time.
From considerations of symmetry, one can show that the most general form
that the polymer contribution to the stress tensor can have in simple
shear flows is~\cite{birda},
\begin{equation}
\btau^p= \pmatrix{
\tau^p_{xx} & \tau^p_{xy} & 0 \cr
\tau^p_{xy} & \tau^p_{yy}& 0 \cr
0 & 0 & \tau^p_{zz} \cr }
\label{sftau}
\end{equation}
where the matrix of components in a Cartesian coordinate system is
displayed.
The form of the stress tensor implies that only three
independent combinations can be measured
for an incompressible fluid. All simple shear flows are consequently
characterised by three material functions. 

\subsubsection{Steady simple shear flows}
Steady simple shear flows are described by a constant shear rate,
${\dot \gamma} = {\vert \dot \gamma_{yx} \vert}$.
The tensor $\bk$ is consequently given by the following matrix
representation in the laboratory-fixed coordinate system,
\begin{equation}
\bk={\dot \gamma} \, \pmatrix{
0 & 1 & 0 \cr
0 & 0 & 0 \cr
0 & 0 & 0 \cr }
\label{ssf1}
\end{equation}
 The three independent
material functions used to characterize such flows are the viscosity,
$\eta_p$, and the first and second normal stress difference coefficients,
$\Psi_1 \,\,{\rm and}\,\, \Psi_2$, respectively.
These functions are defined by the following relations,
\begin{equation}
\tau_{xy}^p = - {\dot \gamma}\, \eta_p \, ; \quad
\tau_{xx}^p- \tau_{yy}^p = - {\dot \gamma^2}\, \Psi_1 \, ; \quad 
\tau_{yy}^p- \tau_{zz}^p = - {\dot \gamma^2}\, \Psi_2 
\label{ssf2}
\end{equation}
where $\tau_{xy}^p, \tau_{xx}^p, \tau_{yy}^p$ are the components of the
polymer contribution to the stress tensor $\btau^p$.

At low shear rates, the viscosity and the first normal
stress coefficient are observed to have constant values, $\eta_{p,0}$
and $\Psi_{1,0},$ termed the zero shear rate
viscosity and the zero shear rate first normal stress coefficient,
respectively. At these shear rates the fluid is consequently  
Newtonian in its behavior.

At higher shear rates, most dilute polymer solutions show 
{\it shear thinning \/} behavior. The viscosity and the first normal
stress coefficient decrease with increasing shear rate, and exhibit a
pronounced {\it power law \/} region. At very high shear rates, the
viscosity has been observed to level off and approach a constant value, 
$\eta_{p,\infty}$, called the infinite shear rate visosity. A high shear
rate limiting value has not been observed for the first normal stress
coefficient. The second normal stress coefficient is much smaller in
magnitude than the first normal stress coefficient, however its sign
has not been conclusively established experimentally. Note that the 
normal stress
differences are zero for a Newtonian fluid. The existence of non-zero
normal stress differences is an indication that the fluid is viscoelastic.

Experiments with very high molecular weight systems seem to suggest that
polymer solutions can also {\it shear thicken}. It has been observed
that the viscosity passes through a minimum with increasing shear rate,
and then increases until a plateau region before shear
thinning again \cite{larson}. 

It is appropriate here to note that shear flow material functions are usually
displayed in terms of the reduced variables, $\eta_p / \eta_{p,0}, \,
\Psi_1 / \Psi_{1,0}$ and $\Psi_2 / \Psi_1$, versus a non-dimensional
shear rate $\beta$, which is defined by $\beta=\lambda_p {\dot \gamma},$ 
where, $\lambda_p = \lbrack \eta \rbrack_0 \, M \, \eta_s / N_{\rm A} \,
k_{\rm B}\, T,$ is a characteristic relaxation time. The subscript 0 on
the square bracket indicates that this quantity is
evaluated in the limit of vanishing shear rate, $k_{\rm B}$ is
Boltzmann's constant and $T$ is the absolute temperature. 
For dilute solutions one can show that, $\lbrack \eta \rbrack /
\lbrack \eta \rbrack_0 = \eta_p / \eta_{p,0} $ and $\beta = \eta_{p,0}
\, {\dot \gamma} / n_p \, k_{\rm B}\, T$. 

\subsubsection{Small amplitude oscillatory shear flow}
A transient experiment that is used very often to characterise polymer
solutions is {\it small amplitude oscillatory shear flow}. The upper
plate in a simple shear experiment is made to undergo 
sinusoidal oscillations in the plane of flow with frequency $\omega$.
For oscillatory flow between narrow slits, 
the shear rate at any position in the fluid is given by \cite{birda},
${ \dot \gamma_{yx} (t)} =
{\dot \gamma_0} \, \cos \omega t$, where ${\dot \gamma_0}$ is the
amplitude. The tensor $\bk(t)$ is consequently given by,
\begin{equation}
\bk(t)={\dot \gamma}_0 \, \cos \, \omega t \pmatrix{
0 & 1 & 0 \cr
0 & 0 & 0 \cr
0 & 0 & 0 \cr } \label{usf3}
\end{equation}

Since the polymer contribution to the shear stress in oscillatory
shear flow, $\tau_{yx}^p$, undergoes a phase shift with respect
to the shear strain and the strain rate, it is customary to
represent its dependence on time through the relation \cite{birda},
\begin{equation}
\tau_{yx}^p=- \eta^\prime(\omega)\, {\dot \gamma}_0 \, \cos \, \omega t
-  \eta^{\prime\prime}(\omega)\,
{\dot \gamma}_0 \, \sin \, \omega t \label{usf4}
\end{equation}
where $\eta^\prime$ and $ \eta^{\prime\prime}$ are the
material functions characterising oscillatory shear flow. It is common
to represent them in a combined form as the complex viscosity,
$\eta^* =\eta^\prime - i \,\eta^{\prime\prime}$.

Two material functions which are entirely equivalent to
$\eta^\prime $ and $\eta^{\prime\prime}$ and which are often used
to display experimental data, are the storage modulus
$G^\prime = \omega \eta^{\prime\prime}/(n k_B T)$ and the loss modulus
$G^{\prime\prime} = \omega  \eta^{\prime}/(n k_B T)$.
Note that the term involving
$G^\prime$ in equation(\ref{usf4}) is in phase with the strain while
that involving $G^{\prime\prime}$ is in phase with the strain rate. For
an elastic material, $G^{\prime\prime}=0$, while for a Newtonian fluid,
$G^\prime=0 $. Thus, $G^{\prime}$ and $G^{\prime\prime}$ are
measures of the extent of the fluid's viscoelasticity. 

In flow situations which have a small displacement gradient, termed the
{\it linear viscoelastic} flow regime, the stress tensor in polymeric
fluids is described by the linear constitutive relation,
\begin{equation}
\btau^p= - \,  \int_{- \infty}^t d\!s \, G(t-s)\,
{\dot \bgam}(t,s)
\label{usf5}
\end{equation}
where $G(t)$ is the relaxation modulus.

When the amplitude $\dot \gamma_0$ is very small,
oscillatory shear flow is a linear viscoelastic flow and consequently
can also be described in terms of a relaxation modulus $G(t)$. Indeed,
expressions for the real and imaginary parts of the complex viscosity
can be found from the expression,
\begin{equation}
\eta^*= \int_0^\infty G(s)\, e^{-i \omega s} \, d\!s
\label{usf6}
\end{equation}

Experimental plots of $\log G^{\prime}$ and $\log G^{\prime\prime}$
versus nondimensional frequency show three distinct power law regimes.
The regime of interest is the intermediate regime \cite {larson}, where
for dilute solutions of high molecular weight polymers in good or theta
solvents, both $G^{\prime}$ and $G^{\prime\prime}$
have been observed to scale with frequency as $\omega^{2/3}$. 

It is appropriate to note here that the zero shear rate viscosity
$\eta_{p,0}$ and the zero shear rate first normal stress
difference $\Psi_{1,0}$, which are linear viscoelastic properties, 
can be obtained from the complex viscosity in the limit of vanishing
frequency,
\begin{equation}
\eta_{p,0} = \lim_{\omega\to 0} \, \eta^{\prime} (\omega) \, ; \quad \quad 
\quad \Psi_{1,0} = \lim_{\omega\to 0} {2 \, \eta^{\prime\prime} 
(\omega) \over \omega} 
\label{usf8}
\end{equation}

\subsection{Scaling with molecular weight}
We have already discussed the scaling of the root mean square end-to-end
distance of a polymer molecule with its molecular weight.
In this section we discuss the scaling of the
zero shear rate intrinsic viscosity
$\lbrack \eta \rbrack_0$, and the translational diffusion coefficient
$D$, with the molecular weight, $M$. As we shall see later, these
have proven to be vitally important as experimental benchmarks in 
attempts to improve predictions of molecular theories. 

It has been found that the relationship between $\lbrack \eta \rbrack_0$
and $M$ can be expressed by the formula,
\begin{equation}
\lbrack \eta \rbrack_0 = K \, M^a
\label{sc1}
\end{equation}
where, $a$ is called the {\it Mark--Houwink \/} exponent, and the
prefactor $K$ depends on the polymer--solvent system. The value of the
parameter $a$ lies between 0.5 and 0.8, with the lower limit
corresponding to theta conditions, and the upper limit to a good solvent
with a very high molecular weight polymer solute. Measured intrinsic
viscosities are routinely used to determine the molecular weight of samples once
the constants $K$ and $a$ are known for a particular polymer--solvent
pair.  

The translational diffusion coefficient  $D$ for a flexible polymer in a
dilute solution can be measured by dynamic light scattering methods, and
is found to scale with molecular weight as \cite{birdb},
\begin{equation}
D \sim M^{-\mu}
\label{sc2}
\end{equation}
where the exponent $\mu$ lies in the range 0.49 to 0.6. Most theta
solutions have values of $\mu$ close to the lower limit. On the other
hand, there is wide variety in the value of $\mu$ reported for good solvents. 
It appears that the upper limit is attained only for very large molecular weight
polymers and the intermediate values, corresponding to a {\it cross over
\/} region, are more typical of real polymers with moderate molecular
weights. 

\subsection{Universal behavior}
It is appropriate at this point to discuss the most important
aspect of the behavior of polymer solutions
(as far as the theoretical modelling of these solutions is concerned)
that is revealed by the various experimental observations. When the 
experimental data for high molecular weight systems is plotted
in terms of appropriately normalized coordinates,
the most noticeable feature is the exhibition of {\it universal \/}
behavior. By this it is meant that curves for different values of a
parameter, such as the molecular weight, the temperature, or even for
different types of monomers can be superposed onto a single curve.
For example, when the reduced intrinsic viscosity,
$\lbrack \eta \rbrack / \lbrack \eta \rbrack_0$
is plotted as a function of the reduced shear rate $\beta$,
the curves for polystyrene in different types of good solvents 
at various temperatures collapse onto a single curve~\cite{birda}.

There is, however, an important point that must be noted. While polymers
dissolved in both theta solvents and good solvents show universal
behavior, the universal behavior is different in the two cases. An example of
this is the observed scaling behavior of various quantities with
molecular weight. The scaling is universal within the context of a particular
type of solvent. The term {\it universality class} is used to describe
the set of systems that exhibit common universal behavior \cite{strobl}.
Thus theta and good solvents belong to different universality classes.

The existence of universality classes is very significant for the
theoretical description of polymer solutions. 
Any attempt made at modelling a polymer solution's properties might expect
that a proper description must incorporate the chemical structure of the
polymer into the model, since this determines its microscopic
behavior. Thus a detailed
consideration of bonds, sidegroups, {\it etc.} may be envisaged. However, the
universal behavior that is revealed by experiments suggests that 
macroscopic properties of the polymer solution are determined by a few large
scale properties of the polymer molecule. Structural details may be
ignored since at length scales in the order of nanometers, different
polymer molecules become equivalent to each other, and behave in the
same manner. As a result, polymer solutions that differ from each
other with regard to the chemical structure or molecular weight
of the polymer molecules that are dissolved in it, the temperature, and
so on, still behave similarly as long as a few parameters that describe
molecular features are the same.    

This universal behavior justifies the introduction of crude mechanical
models, such as the bead-spring chain, to represent real polymer molecules.
On the other hand, it is interesting to note that in many 
cases, the predictions of these models are not universal. It turns
out that apart from a basic length and time scale,
there occur other parameters that need to be prescribed, for example, the
number of beads $N$ in the chain, the strength of hydrodynamic
interaction $h^*$, the finite spring extensibility parameter $b$, and so on. 
It is perhaps not incorrect to state that any molecular theory that is
developed must ultimately verify that universal predictions
of transport properties are indeed obtained. The universal 
predictions of kinetic theory models with hydrodynamic 
interaction are discussed later on in this chapter. 

\section{BEAD-SPRING CHAIN MODELS}
\label{bscm}
The development of a kinetic theory for dilute solutions has been approached
in two different ways. One of them is an intuitive approach in the
configuration space of a single molecule, with a particular mechanical
model chosen to represent the macromolecule, such as a freely rotating
bead-rod chain or a freely jointed bead-spring chain \cite{kirkwood,rouse,zimm}.
The other approach is to develop a formal theory in
the phase space of the entire solution, with the polymer molecule
represented by a general mechanical model that may have internal constraints,
such as constant bond lengths and angles \cite{kramers,cbh,birdb}.
The results of the former
method are completely contained within the latter method, and several ad hoc
assumptions made in the intuitive treatment are clarified and placed in
proper context by the development of the rigorous phase space theory.
Kinetic theories developed for {\it flexible} macromolecules in dilute
solutions have generally pursued the intuitive approach, with the
bead-spring model proving to be the most popular. This is because the lack of
internal constaints in the model makes the formulation of the theory simpler.
Recently, Curtiss and Bird \cite{cb}, acknowledging the
`notational and mathematical' complexity of the rigorous phase space
theory for general mechanical models, have summarised the results of
phase space theory for the special case of bead-spring models with
arbitrary connectivity, {\it ie.} for linear chains, rings, stars, combs and
branched chains.

In this section, since we are primarily concerned with reviewing recent
developments in theories for flexible macromolecules, we describe the
development of kinetic theories in the configuration space of a single molecule.
However, readers who wish the understand the origin of the ad hoc
expressions used for the Brownian forces and the hydrodynamic force,
and the formal development of expressions for the momentum and mass
flux, are urged to read the article by Curtiss and Bird \cite{cb}. 

The general diffusion equation that governs the time evolution of
the distribution of configurations of a bead-spring chain subject to
various nonlinear effects, and the microscopic origin of the polymer
contribution to the stress tensor are discussed in this section. 
The simplest bead-spring chain model, the {\it Rouse} model is also
discussed. We begin, however, by describing the equilibrium 
statistical mechanical arguments that justify the
representation of a polymer molecule with a bead-spring chain model,
and we discuss the equilibrium configurations of such a model. 

\subsection{Equilibrium configurations}
When a flexible polymer chain in a {\it quiescent} dilute
solution is considered at a lowered resolution,
{\it ie.} at a coarse-grained level, it would appear like a
strand of highly coiled spaghetti, and
the extent of its coiling would depend on its degree of flexibility. A
quantity used to characterise a chain's flexibility is the
{\it orientational correlation function}, whose value $K_{\rm or}\, (\Delta
\ell)$, is a measure of the correlation in the direction of the chain at
two different points on the chain which are
separated by a distance $\Delta \ell$ along the
length of the chain. At sufficiently large distances
$\Delta \ell$, it is expected that the correlations vanish. However, it is
possible to define a {\it persistence length \/} $\ell_{\rm ps}$, such
that for $\Delta \ell > \ell_{\rm ps}$, orientational
correlations are negligible \cite{strobl}. 

The existence of a persistence length suggests that as far as the
global properties of a flexible polymer chain are concerned,
such as the distribution function for the end-to-end
distance of the chain, the continuous chain could be replaced by
a freely jointed chain made up of
rigid links connected together at joints that are 
completely flexible, whose linear segments are each longer than
the persistence length $\ell_{\rm ps}$, and whose contour length is the
same as that of the continuous chain.

The freely jointed chain undergoing thermal motion is clearly
analogous to a random-walk in space, with each random step in the walk
representing a link in the chain assuming a random orientation. Thus all the
statistical properties of a random-walk are, by analogy,
also the statistical properties of the freely jointed chain.  The
equivalence of a polymer chain with a random-walk lies at the heart
of a number of fundamental results in polymer physics. 

\subsubsection{Distribution functions and averages}
In polymer kinetic theory, the freely jointed chain is assumed to have
beads at the junction points betwen the links,
and is referred to as the freely jointed bead-rod chain \cite{birdb}.
The introduction of the beads is to account for the mass of the polymer
molecule and the viscous drag experienced by the polymer molecule.
While in reality the mass and drag are distributed continuously
along the length of the chain, the model assumes
that the total mass and drag may be distributed over a finite number of
discrete beads. 

For a general chain model consisting of $N$ beads, which have  
position vectors ${\br}_{\nu}, \, \nu = 1,2, \ldots, N,$ in a 
laboratory fixed coordinate system, the Hamiltonian is given by,
\begin{equation}
{\cal H} = {\cal K} +
\phi \, ({\br}_1,{\br}_2, \ldots, {\br}_N)
\label{ham}
\end{equation}
where $\cal K$ is the kinetic energy of the system and $\phi$ is the
potential energy. $\phi$ depends on the location of all the
particles. 

The center of mass ${\br}_c$ of the chain, and its velocity 
${\dot {\br}}_c$ are given by
\begin{equation}
{\br}_c = {1 \over N }\, \sum_{\nu=1}^N \, {\br}_{\nu}  \quad ; \quad 
{\dot {\br}}_c = {1 \over N }\, \sum_{\nu=1}^N \, {\dot {\br}}_{\nu}
\end{equation}
where ${\dot {\br}}_{\nu} = d {\br}_{\nu}/dt$.
The location of a bead with respect to the center of mass is specified
by the vector ${\bR}_{\nu} = {\br}_{\nu} - {\br}_c$.

If $Q_1, \, Q_2, \, \ldots \, Q_d$ denote the generalised internal
coordinates required to specify the configuration of the
chain, then the kinetic energy of the chain in terms
of the velocity of the center of mass and the generalised velocities
${\dot Q}_s = d Q_s /dt$, is given by~\cite{birdb},
\begin{equation}
{\cal K} = {m \,N \over 2 } \, {\dot {\br}}_c^2 + 
{1 \over 2} \, \sum_s \sum_t \,
g_{st}\, {\dot Q}_s \,{\dot Q}_t
\end{equation}
where the indices $s$ and $t$ vary from 1 to $d$,
$m$ is the mass of a bead, and 
$g_{st}$ is the {\it metric matrix}, defined by, 
$g_{st} = m \, \sum_{\nu}\, ({\partial  {\bR}_{\nu} / \partial Q_s}) 
\cdot ({\partial  {\bR}_{\nu} / \partial Q_t} ) $. 
In terms of the momentum of the center of mass,
${\bp}_c = m \, N \, {\dot \br}_c$, and the generalised momenta
$P_s$, defined by,
$P_s = ({\partial  {\cal K} / \partial {\dot Q}_s}), $
the kinetic energy has the form~\cite{birdb},  
\begin{equation}
{\cal K} = {1 \over 2 m \,N} \, {\bp}_c^2 + {1 \over 2} \sum_s \sum_t
G_{st}\,  P_s \,P_t
\label{kin}
\end{equation}
where, $G_{st}$ are the components of the matrix inverse to the metric matrix,
$\sum_t \, G_{st} \, g_{tu} = \delta_{su},$ and  $\delta_{su}$ is the 
Kronecker delta.

The probability, ${\cal P}_{\rm eq} \, d{\br}_c \, dQ \, d{\bp}_c \,d P$,
that an $N$-bead chain model has a configuration
in the range $ \, d{\br}_c \, dQ \,$ about $ \, {\br}_c, \, Q \,$ and
momenta in the range 
$\, d{\bp}_c \, dP \,$ about $ \, {\bp}_c, \, P \,$ is given by,
\begin{equation}
{\cal P}_{\rm eq}\, \bigl( \, {\br}_c, \, Q, \, {\bp}_c, \, P \, \bigr)
= {\cal Z}^{-1} \, e^{- {\cal H} / k_{\rm B} T}
\end{equation}
where $\cal Z$ is the {\it partition function}, defined by,
\begin{equation}
{\cal Z} = \int\!\!\int\!\!\int\!\!\int \, e^{- {\cal H} / k_{\rm B}T} \, 
d{\br}_c \, dQ \, d{\bp}_c \, dP
\end{equation}
The abbreviations, $Q$ and $dQ$ have been used to denote $\, Q_1, \,
Q_2, \, \ldots, \, Q_d \, $ and $\, dQ_1 \, dQ_2 \, \ldots \,
dQ_d \,$, respectively, and a similar notation has been used for the momenta. 

The {\it configurational distribution function} for a general $N$-bead chain,
$\psi_{\rm eq} \, (\, Q \, ) \, dQ$,
which gives the probability that the internal configuration
is in the range $\, dQ\, $ about $\, Q \, $, 
is obtained by integrating
${\cal P}_{\rm eq}$ over all the momenta and over the coordinates of
the center of mass,
\begin{equation}
\psi_{\rm eq} \, ( \,  Q \, )
= {\cal Z}^{-1} \,  {\int\!\!\int\!\!\int \, e^{- {\cal H} / k_{\rm B} T}
d{\br}_c \, d{\bp}_c \, dP }
\end{equation}
For an $N$-bead chain whose potential energy does not depend on the
location of the center of mass, the following result is obtained 
by carrying out the
integrations over ${\bp}_c$ and $P$ \cite{birdb},
\begin{equation}
\psi_{\rm eq} \, ( \,  Q \, )
= {\sqrt{g(Q)} \, e^{- {\phi(Q)} / k_{\rm B} T} \over
\int {\sqrt{g(Q)} \, e^{- {\phi(Q)} / k_{\rm B} T} \, dQ }}
\label{confdis}
\end{equation}
where, $g(Q)=\det (g_{st}) = 1/\det (G_{st})$.

An expression that is slightly different from the random-walk distribution
is obtained on evaluating the right hand side of equation~(\ref{confdis})
for a freely jointed bead-rod chain.
Note that the random-walk distribution is obtained by assuming that
each link in the chain is oriented independently of all the other links,
and that all orientations of the link are equally likely.
On the other hand, equation~(\ref{confdis}) suggests that
the probability for the links in a freely jointed chain
being perpendicular to each other, for a given solid angle, is slightly
larger than the probability of being in the same direction. 
Inspite of this result, the configurational
distribution function for a freely jointed bead-rod chain is almost
always assumed to be given by the random-walk distribution \cite{birdb}.
Here afterwards in this chapter, we shall refer to a freely
jointed bead-rod chain whose configurational distribution function is
assumed to be given by the random-walk distribution, as an {\it ideal } chain. 
For future reference, note that the ran\-dom-walk dis\-tribution is given by,
\begin{equation}
\psi_{\rm eq} \, ( \,\theta_1, \ldots, \theta_{N-1},
\phi_1, \ldots, \phi_{N-1}  \, ) = \Biggl( {1 \over 4 \pi} \Biggr)^{N-1}
 \, \prod_{i=1}^{N-1} \, \sin \, \theta_i 
\label{ranwalk}
\end{equation}
where $\theta_i$ and $\phi_i$ are the polar angles for the $i {\rm th}$
link in the chain \cite {birdb}.

Since the polymer chain explores many states in the duration of an observation 
quantities observed on macroscopic length and time scales 
are {\it averages} of functions of the configurations and
momenta of the polymer chain. A few definitions of averages are 
now introduced that are used frequently subsequently in the chapter.

The average value of a function
$X \, \bigl( \, {\br}_c, \, Q, \, {\bp}_c, \, P \, \bigr)$, defined in
the phase space of a polymer molecule is given by, 
\begin{equation}
\avel X \aver_{\rm eq} = \int\!\!\int\!\!\int\!\!\int \,
X \, {\cal P}_{\rm eq}\, d{\br}_c \, dQ \, d{\bp}_c \, dP
\end{equation}
We often encounter quantities $X$ that depend only on the internal
configurations of the polymer chain and not on the center of mass
coordinates or momenta. In addition, if the potential energy of the
chain does not depend on the location of the center of mass, then it is
straight forward to see that the equilibrium average of X is given by,
\begin{equation}
\avel X \aver_{\rm eq} = \int \, X \, \psi_{\rm eq} \, dQ 
\label{conave}
\end{equation}

\subsubsection{The end-to-end vector}
 The end-to-end vector ${\br}$ of a general bead-rod chain can be
found by summing the vectors that represent each link in the chain,
\begin{equation}
{\br} = \sum_{i=1}^{N-1} \, a \, {\bfu}_i
\end{equation}
where $a$ is the length of a rod, and ${\bfu}_i$ is a unit vector in
the direction of the $i \rm{th}$ link of the chain. Note that the components of
the unit vectors ${\bfu}_i, \, i=1,2,\ldots,N-1 \,$, can be 
expressed in terms of the generalised coordinates $Q$ \cite{birdb}.

The probability $P_{\rm eq} ({\br}) \, d {\br}$, that the end-to-end vector
of a general bead-rod chain is in the range $d {\br}$
about ${\br}$ can be found
by suitably contracting the configurational distribution function
$\psi_{\rm eq} \, ( \, Q \, )$ \cite{birdb},
\begin{equation}
P_{\rm eq} ({\br}) 
= \int  \delta \Biggl(\, {\br} - \sum_i \,  a \, {\bfu}_i \, \Biggr) \,
\psi_{\rm eq} \, ( \, Q \, ) \, dQ
\label{endvec}
\end{equation}
where $\delta ( . )$ represents a Dirac delta function. 

With $\psi_{\rm eq} \, ( \, Q \, )$ given by the ran\-dom-walk
distribution~(\ref{ranwalk}), it can be shown that for large values
of $N$ and $r = \vert {\br} \vert < 0.5 N a $, the probability
distribution for the end-to-end vector is a Gaussian distribution,
\begin{equation}
P_{\rm eq} ({\br}) 
= \Biggl( \, {3 \over 2  \pi  (N-1) a^2 } \, \Biggr)^{3/2} \,
\exp \, \Biggl({-3 \, r^2 \over 2 \, (N-1) \, a^2 } \Biggr) 
\label{gendvec}
\end{equation}
The distribution function for the end-to-end vector of an {\it ideal} chain 
with a large number of beads $N$ is therefore given by the Gaussian
distribution~(\ref{gendvec}).

The mean square end-to-end distance, $\avel \, r^2 \, \aver_{\rm eq},$  
for an ideal chain can then be shown to be, 
$\avel \, r^2 \, \aver_{\rm eq} = ( N-1 ) \, a^2.$
This is the well known result that the root mean square of the
end-to-end distance of a random-walk increases as the square root of the
number of steps. In the context of the polymer chain, since the
number of beads in the chain is directly proportional to the molecular
weight, this result implies that $R \sim M^{0.5}$. We have seen earlier
that this is exactly the scaling observed in theta solvents. Thus one
can conclude that a polymer chain in a theta solvent behaves like an
ideal chain.

\subsubsection{The bead-spring chain}
Consider an isothermal system consisting of a bead-rod chain with a
constant end-to-end vector ${\br}$, sus\-pended in a bath of so\-lvent
molecules at temperature $T$. 
The partition function of such a constrained system can be 
found by contracting
the partition function in the con\-straint-free case, 
\begin{equation}
{\cal Z} \, ({\br}) = \int\!\!\int\!\!\int\!\!\int \, 
\delta \Biggl(\, {\br} - \sum_i \,  a \, {\bfu}_i \, \Biggr) \,
e^{- {\cal H} / k_{\rm B}T} \,
d{\br}_c \, dQ \, d{\bp}_c \, dP
\label{parfun}
\end{equation}
For an $N$-bead chain whose potential energy does not depend on the
location of the center of mass, the integrations over ${\br}_c$,
${\bp}_c$ and $P$ can be carried out to give,
\begin{equation}
{\cal Z} \, ({\br}) = C  \, \int \, 
\delta \Biggl(\, {\br} - \sum_i \,  a \, {\bfu}_i \, \Biggr) \,
\psi_{\rm eq} \, ( \,  Q \, )
d{\br}_c \, dQ \end{equation}
Comparing this equation with the equation for the end-to-end
vector (\ref{endvec}), one can conclude that, 
\begin{equation}
{\cal Z} \, ({\br}) = C  \, P_{\rm eq} \, ({\br})
\label{constraintpf}
\end{equation}
In other words, the partition function of a general bead-rod chain
(except for a multiplicative factor independent of ${\br}$)
is given by $P_{\rm eq} \, ({\br})$. This result
is essential to establish the motivation for the introduction of the
bead-spring chain model. 

At constant temperature, the change in free energy accompanying a change
in the end-to-end vector ${\br}$ of a bead-rod chain, by an infinitesimal amount
$d{\br}$, is equal to the work done in the process,
{\it ie.,} $dA = {\bF} \cdot d{\br}$, where ${\bF}$ is the force
required for the extension.
The Helmholtz free energy of a general bead-rod chain with fixed end-to-end
vector ${\br}$ can be found from equation~(\ref{constraintpf}), 
\begin{equation}
A( {\br})= - \, k_{\rm B} T \, \ln {\cal Z} ({\br}) 
=  A_0 - k_{\rm B} T \, \ln   P_{\rm eq} \, ({\br})
\label{helm}
\end{equation}
where $A_0$ is a constant independent of ${\br}$.
For an ideal chain, it follows from
equations (\ref{gendvec}) and (\ref{helm}), that a change in the
end-to-end vector 
by $d{\br}$, leads to a change in the free energy $dA$, given by, 
\begin{equation}
dA( {\br}) = {3 k_{\rm B} T \over (N-1) a^2 } \; {\br} \cdot d{\br} 
\label{spring}
\end{equation}
Equation (\ref{spring}) implies that there is a {\it tension} 
${\bF}$ in the ideal chain,
$ {\bF} = ( {3 k_{\rm B} T / (N-1) a^2 } ) \, {\br}, $
which resists any attempt at chain extension. Furthermore, this tension is
proportional to the end-to-end vector ${\br}$. This implies that the
ideal chain acts like a {\it Hookean} spring, with a spring constant $H$
given by,
\begin{equation}
H =  {3 k_{\rm B} T \over (N-1) a^2 }
\end{equation}

The equivalence of the behavior of an ideal chain to that of a Hookean spring
is responsible for the introduction of
the bead-spring chain model. Since long enough {\it
sub-chains} within the ideal chain also have normally distributed
end-to-end vectors, the entire ideal chain may be replaced by beads
connected to each other by springs. Note that each bead in a bead-spring chain
represents the mass of a sub-chain of the ideal chain, while the spring
imitates the behavior of the end-to-end vector of the sub-chain. 

The change in the Helmholtz free energy of an ideal chain due to a
change in the end-to-end vector is purely due to entropic
considerations. The internal energy, which has only the kinetic 
energy contribution, does not depend on the end-to-end vector. 
Increasing the end-to-end vector of the chain decreases the
number of allowed configurations, and this change is resisted by the
chain. The entropic origin of the resistance is responsible for the
use of the phrase {\it entropic spring} to describe the springs of the
bead-spring chain model.

The potential energy $S$, of a bead-spring chain due to the presence of
Hookean springs is the sum of the
potential energies of all the springs in the chain. For a bead-spring
chain with $N$ beads, this is given by,
\begin{equation}
S = {1 \over 2} \, H \, \sum_{i=1}^{N-1} \, {\bQ}_i \cdot {\bQ}_i
\end{equation}
where ${\bQ}_i= {\br}_{i+1} - {\br}_{i}$
is the {\it bead connector vector} between the beads $i$ and $i+1$.
The configurational distribution function for a Hookean bead-spring chain may be
found from equation~(\ref{confdis}) by substituting $\phi(Q) = S$,
with the Cartesian components of the connector
vectors chosen as the generalised coordinates $Q_s$. The number of
generalised coordinates is consequently, $d=3N-3$, reflecting the lack of any
constraints in the model. Since $g(Q)$ is a constant independent of $Q$ for
the bead-spring chain model \cite{birdb}, one can show that,
\begin{equation}
\psi_{\rm eq} \, (\,{\bQ}_1, \ldots, \, {\bQ}_{N-1} \,)
= \prod_j \, \Biggl( \, {H \over 2  \pi  k_{\rm B} T } \, \Biggr)^{3/2} \,
\exp \, \Biggl( {- H \over 2 \,k_{\rm B} T } \; {\bQ}_j \cdot {\bQ}_j \Biggr)   
\label{equidis}
\end{equation}
It is clear from equation~(\ref{equidis}) that the equilibrium
distribution function for each connector vector
in the bead-spring chain is a Gaussian distribution, and these
distributions are independent of each other. From the property of
Gaussian distributions, it follows that the vector connecting any
two beads in a bead-spring chain at equilibrium also obeys a Gaussian
distribution. 

The Hookean bead-spring chain model has the unrealistic feature that the
magnitude of the end-to-end vector has no upper bound and can infact
extend to infinity. On the other hand, the real polymer molecule
has a finite fully extended length. This deficiency of the bead-spring
chain model is not serious at equilibrium, but becomes important in
strong flows where the polymer molecule is highly extended. Improved models 
seek to correct this deficiency by modifying the force law between the beads of
the chain such that the chain stiffens as its extension increases. An 
example of such a nonlinear spring force law that is very commonly 
used in polymer literature is the {\it finitely extensible 
nonlinear elastic} (FENE) force law~\cite{birdb}. 

\subsubsection{Excluded volume}
The universal behavior of polymers dissolved in theta solvents can be
explained by recognising that all high molecular weight polymers 
dissolved in theta solvents behave like ideal chains.  However, a polymer chain
cannot be identical to an ideal chain since unlike the ideal chain,
two parts of a polymer chain cannot occupy the same location at the
same time. In the very special case of a theta solvent, the excluded
volume force is just balanced by the repulsion of the solvent molcules.
In the more commonly occuring case of good solvents, the excluded volume
interaction acts between any two parts of the chain
that are close to each other in space, irrespective of their distance
from each other along the chain length, and leads to a swelling of the chain.
This is a {\it long range interaction}, and as a result,
it seriously alters the macroscopic
properties of the chain. Indeed there is a qualitative difference, and
this difference cannot be treated as a small perturbation from the
behavior of an ideal chain~\cite{strobl}. Curiously enough
however, all swollen chains behave similarly to each other, and 
modelling this universal behavior was historically one of the challenges of 
polymer physics~\cite{strobl,yamakawa,degennes,doied,desclo}. 
Here, we very briefly mention the manner in which the problem is 
formulated in the case of bead-spring chains. 

The presence of excluded volume causes the polymer chain to swell.
However, the swelling ceases when the
entropic retractive force balances the excluded
volume force. The retractive force arises due
to the decreasing number of conformational states available to the
polymer chain due to chain expansion. This picture of the microscopic 
phenomenon is captured by writing the potential energy of the 
bead-spring chain as a sum of the spring potential energy and 
the potential energy due to excluded volume interactions. 
The excluded volume potential energy is found by summing the 
interaction energy over all pairs of beads $\mu$ and $\nu$,
$E = (1 / 2) \sum_{\mu,\nu = 1 \atop \mu \ne \nu}^N \, E \left( {\br}_{\nu}
- {\br}_{\mu} \right), $
where $E \left( {\br}_{\nu} - {\br}_{\mu} \right)$ is a short-range
function usually taken as,
$ E \left( {\br}_{\nu} - {\br}_{\mu} \right) = v \, k_{\rm B} T \, 
\delta \left( {\br}_{\nu} - {\br}_{\mu} \right);$
$v$ being the excluded volume parameter with dimensions of volume. 
The total potential energy of a Hookean bead-spring chain with 
$\delta$-function excluded volume interactions is consequently,
\begin{equation}
\phi = {1 \over 2} \, H \, \sum_{i=1}^{N-1} \, {\bQ}_i \cdot {\bQ}_i 
+ {1 \over 2} \, v \, k_{\rm B} T \, \sum_{\mu,\nu = 1 \atop \mu \ne \nu}^N \,
\delta \left( {\br}_{\nu} - {\br}_{\mu} \right) 
\label{phitot}
\end{equation}

The equilibrium configurational distribution function of a polymer chain
in the presence of Hookean springs and excluded volume can be
found by substituting equation~(\ref{phitot}) into
equation~(\ref{confdis}), and all average properties of the chain can be
found by using equation~(\ref{conave}). Solutions to these equations in
the limit of long chains have been found by using a number of
approximate schemes since an exact treatment is impossible. The most
accurate scheme involves the use of field theoretic and renormalisation group
methods~\cite{desclo}. The universal scaling of a 
number of equilibrium properties of
dilute polymer solutions with good solvents are correctly predicted
by this theory. For instance, the end-to-end distance is predicted to scale
with molecular weight as, $R \sim M^{0.588}$. 

The spring potential in equation~(\ref{phitot}) has been
derived by considering the Helmholtz free energy of an ideal chain, 
{\it ie.} under theta conditions. It seems reasonable to expect 
that a more accurate derivation
of the retractive force in the chain due to entropic considerations
would require the treatment of a polymer chain
in a good solvent. This would lead to a
non-Hookean force law between the beads~\cite{degennes,ottbook}. 
Such non-Hookean force laws have so far not been treated in
non-equilibrium theories for dilute polymer solutions with good
solvents.

\subsection{Non-equilibrium configurations}
Unlike in the case of equilibrium solutions   
it is not possible to derive the phase space distribution function 
for non-equilibrium solutions from very general arguments. 
As we shall see here it is only possible to derive a partial 
differential equation that governs the evolution of the configurational 
distribution function by considering the conservation of  
probability in phase space, and the equation of motion for
the particular model chosen. The arguments relevent to a bead-spring chain 
are developed below. 

\subsubsection{Distribution functions and averages}
The phase space of a bead-spring chain with $N$ beads can be chosen to
be given by the $6N - 6$ components of the bead position coordinates,
and the bead velocities such that,
\[ {\cal P} \, \bigl( \, {\br}_1, \ldots, {\br}_N, \,
{ {\dot \br}}_1, \ldots, { {\dot \br}}_N, \, t \, \bigr) \, 
d{\br}_1 \ldots d{\br}_N \, 
d { {\dot \br}}_1 \ldots d{ {\dot \br}}_N \]
is the probability that the bead-spring chain has an instantaneous
configuration in the range $d{\br}_1, \ldots , d{\br}_N$ about
$ {\br}_1, \ldots , {\br}_N$, and the beads in the chain
have velocities in the range 
$d { {\dot \br}}_1, \ldots , d{ {\dot \br}}_N$ about
$ { {\dot \br}}_1, \ldots , { {\dot \br}}_N$.

The configurational distribution function $\Psi$,
can be found by integrating ${\cal P}$ over all the bead velocities,
\begin{equation}
\Psi \, (  \, {\br}_1, \ldots, \, {\br}_{N}, \, t \,)  
= \int\! \ldots \!\int \, {\cal P} \, 
d { {\dot \br}}_1 \ldots d{ {\dot \br}}_N
\end{equation}
The distribution of internal configurations $\psi $, is given by,
\begin{equation}
\psi \, ( \, {\bQ}_1, \ldots, \, {\bQ}_{N-1}, \, t \,)  
=  \int \, \Psi^\prime \, ( \, {\br}_c, \, {\bQ}_1, 
\ldots, \, {\bQ}_{N-1}, \, t \,) \, d {\br}_c
\end{equation}
where, $ \Psi^\prime = \Psi,$ as a result of the 
Jacobian relation for the configurational vectors~\cite{birdb}, 
\[ 
\left\vert {\, {\partial( \, {\br}_1, \ldots, \, {\br}_{N} \, )  \over 
\partial ( \, {\br}_c, {\bQ}_1, \ldots, \, {\bQ}_{N-1} \,)} }\, 
\right\vert = 1
\] 
Note that the normalisation condition
$\int \, \psi \, d{\bQ}_1 \, d{\bQ}_2 \, \ldots \, d{\bQ}_{N-1}
= 1$ is satisfied by $\psi$. 
When the configurations of the bead-spring chain do not depend on the location
of the center of mass, as in the case of homogeneous flows with no concentration
gradients, $ (1 / V) \, \psi  = \Psi, $ 
where $V$ is the volume of the solution.

The velocity-space distribution function $\Xi$ is defined by,
\begin{equation}
{\Xi } \, \bigl( \, {\br}_1, \ldots, {\br}_N, \,
{ {\dot \br}}_1, \ldots, { {\dot \br}}_N, \, t \, \bigr) \, 
= {{\cal P} \over \Psi}
\end{equation}
Note that $\Xi$ satisfies the normalisation condition
$\int \! \ldots \! \int  \, \Xi \, 
d { {\dot \br}}_1 \ldots d{ {\dot \br}}_N = 1$. 
Under certain circumstances that are discussed later, it is common to
assume that the velocity-space distribution function is
Maxwellian about the mass-average solution velocity,
\begin{equation}
\Xi = {\cal N}_M \,
\exp \, \biggl[ \, - \, {1 \over 2 \, k_{\rm B} T } \,
\Bigl[ \,  m ( { {\dot \br}}_1 -
{\bv})^2  + \ldots + m ( { {\dot \br}}_N -
{\bv})^2 \, \Bigr] \,   \biggr] 
\end{equation}
where ${\cal N}_M$ is the normalisation constant for the Maxwellian
distribution. Making this assumption implies that one expects
the time scales involved in equilibration processes in
momentum space to be much smaller than
the time scales governing relaxation processes in configuration space. 

Averages of quantities which are functions of the bead positions and 
bead velocities are defined analogously to the those in the
previous section, namely,
\begin{equation}
\avel X \aver = \int \! \ldots \! \int \,
X \, {\cal P} \, d{\br}_1 \ldots d{\br}_N \, 
d { {\dot \br}}_1 \ldots d{ {\dot \br}}_N 
\end{equation}
is the the phase space average of $X$, while the velocity-space average is,
\begin{equation}
\lmav  X \, \rmav
= \int\!\!\int \, X \, \Xi \, d { {\dot \br}}_1 \ldots d{ {\dot \br}}_N 
\end{equation}
For quantities $X$ that depend only on the internal
configurations of the polymer chain and not on the center of mass
coordinates or bead velocities, 
\begin{equation}
\avel X \aver = \int \, X \, \psi \, d{\bQ}_1 d{\bQ}_2 
\ldots d{\bQ}_{N-1} 
\end{equation}

\subsubsection{The equation of motion}
The equation of motion for a bead in a bead-spring chain is
derived by considering the forces acting on it. The
total force ${\bF}_{\mu}$, on bead $\mu$ is,
${\bF}_{\mu} = \sum_i \, {\bF}_{\mu}^{(i)}$,
where the ${\bF}_{\mu}^{(i)}, \; i =1,2, \ldots$, are the various intramolecular
and solvent forces acting on the bead.
The fundamental difference among the
various molecular theories developed so far for the description
of dilute polymer solutions lies in the
kinds of forces ${\bF}_{\mu}^{(i)}$ that are assumed to be acting on
the beads of the chain. In almost all these theories, the accelaration of
the beads due to the force ${\bF}_{\mu}$ is neglected.
A bead-spring chain model incorporating bead inertia has shown that
the neglect of bead inertia is justified in most practical
situations~\cite{schiebott}. The equation of motion is consequently obtained by
setting ${\bF}_{\mu} = {\bzero}$.
Here, we consider the following force balance on each bead $\mu$,
\begin{equation}
{\bF}_{\mu}^{(h)} + {\bF}_{\mu}^{(b)} + {\bF}_{\mu}^{(\phi)} + 
{\bF}_{\mu}^{(iv)} = {\bzero} \quad (\mu = 1, 2, \ldots, N)
\label{eqmo}
\end{equation}
where, ${\bF}_{\mu}^{(h)}$ is the {\it hydrodynamic drag } force,
${\bF}_{\mu}^{(b)}$ is the {\it Brownian } force, 
${\bF}_{\mu}^{(\phi)}$ is the {\it intramolecular } force due to
the potential energy of the chain, and
${\bF}_{\mu}^{(iv)}$ is the force due to the presence of
{\it internal viscosity}. These are the various forces
that have been considered so far in the literature, which
are believed to play a crucial role in determining the polymer solution's
transport properties.
The nature of each of these forces is discussed in greater detail below.
Note that, as is common in most theories, external forces acting on the
bead have been neglected. However, their inclusion is reasonably straight 
forward~\cite{birdb}.  

The hydrodynamic drag force ${\bF}_{\mu}^{(h)}$,
is the force of resistance offered by the
solvent to the motion of the bead $\mu$. It is assumed to be proportional to
the difference between the velocity-averaged bead velocity
$\lmav { {\dot \br}}_{\mu} \rmav$ and the local velocity of the solution,
\begin{equation}
{\bF}_{\mu}^{(h)} = - \zeta \, [ \, \lmav { {\dot \br}}_{\mu} \rmav
 - ({\bv}_{\mu} + {\bv}^\prime_{\mu}) \, ]
\end{equation}
where $\zeta$ is bead friction coefficient.
Note that for spherical beads with radius $a$, in a solvent with viscosity
$\eta_s$, the bead friction coefficient $\zeta$ is given by the Stokes
expression: $\zeta=6 \pi \eta_s a$. The velocity-average of the
bead velocity is not carried out with the Maxwellian distribution since
this is just the mass-average solution velocity. However, it turns out that
an explicit evaluation of the velocity-average is unnecessary for the
development of the theory. Note that 
the velocity of the solution at bead $\mu$ has two components, 
the imposed flow field ${\bv}_{\mu} = \bv_0 + \bk (t) \cdot 
{\br}_{\mu}$, and the perturbation of the flow field
${\bv}^\prime_{\mu}$ due to the motion of the other beads of the
chain. This perturbation is called {\it hydrodynamic interaction}, and its
incorporation in molecular theories has proved to be of utmost
importance in the prediction of transport properties.
The presence of hydrodynamic interaction couples the motion of one bead
in the chain to all the other beads, regardless of the distance between
the beads along the length of the chain. In this sense, hydrodynamic
interaction is a long range phenomena. 

The perturbation to the flow field ${\bv}^\prime ({\br})$ at a
point ${\br}$ due to the presence of a point force ${\bF} 
({\br}^\prime)$ at the point ${\br}^\prime$, can be found by solving the
linearised Navier-Stokes equation~\cite{birda,doied},
\begin{equation}
{\bv}^\prime ({\br}) = \bO ({\br} - {\br}^\prime) \cdot 
{\bF} ({\br}^\prime)
\end{equation}
where $\bO ({\br})$, called the {\it Oseen-Burgers tensor}, is the
Green's function of the linearised Navier-Stokes equation,
\begin{equation}
\bO ({\br}) = {1 \over 8 \pi \eta_s r} \, 
\biggl( \bu + {\br \br \over r^2} \biggr)
\end{equation}

The effect of hydrodynamic interaction is taken into account in polymer
kinetic theory by treating the beads in the bead-spring chain as point
particles. As a result, in response to the hydrodynamic drag force
acting on each bead, each bead exerts an equal and opposite force on
the solvent at the point that defines its location. The disturbance to
the velocity at the bead $\nu$ is the sum of the disturbances caused by
all the other beads in the chain, 
${\bv}^\prime_{\nu} = - \sum_{\mu} \,
\bO_{\nu \mu} ({\br_{\nu}-\br_{\mu}}) \cdot {\bF}_{\mu}^{(h)}, $
where, $\bO_{\mu \nu} = \bO_{\nu \mu} $ is given by,
\begin{equation}
\bO_{\mu \nu}= \cases{\displaystyle{{1 \over
8 \pi \eta_s r_{\mu \nu}}}
\biggl( \displaystyle{\bu + {\br_{\mu \nu} \br_{\mu \nu} \over
r_{\mu \nu}^2}} 
\biggr), \quad \br_{\mu \nu} =
\br_{\mu}-\br_{\nu}, & for $\mu \neq \nu$ \cr
\noalign{\vskip3pt}
0 & for $\mu = \nu$ \cr}
\end{equation}

The Brownian force ${\bF}_{\mu}^{(b)}$, on a bead $\mu$ is the
result of the irregular collisions between the solvent molecules and the
bead. Instead of representing the Brownian force by a randomly varying
force, it is common in polymer kinetic theory to use an averaged
Brownian force,
\begin{equation}
{\bF}_{\mu}^{(b)} = - k_{\rm B} T \, \biggl( \, {\partial \, \ln \Psi \over
\partial {\br}_{\mu}} \, \biggr)
\label{browf}
\end{equation}
As mentioned earlier, the origin of this expression can be understood
within the framework of the complete phase space theory~\cite{birdb,cb}.
Note that the Maxwellian distribution has been used to
derive equation~(\ref{browf}). 

The total potential energy $\phi$ of the bead-spring chain is the sum of the
potential energy $S$ of the elastic springs, and the potential energy
$E$ due to the presence of excluded volume interactions between the beads.
The force ${\bF}_{\mu}^{(\phi)}$ on a bead $\mu$ due to the intramolecular
potential energy $\phi$ is given by,
\begin{equation}
{\bF}_{\mu}^{(\phi)} = -  {\partial \phi \over
\partial {\br}_{\mu}} 
\label{potf}
\end{equation}

In addition to the various forces discussed above,
the {\it internal viscosity} force ${\bF}_{\mu}^{(iv)}$, has received
considerable attention in literature~\cite{birdott,schiebiv,wedgeiv}, 
though it appears not to have widespread acceptance. Various physical 
reasons have been cited as 
being responsible for the internal viscosity force. For instance, the 
hindrance to internal rotations due to the presence of energy barriers,
the friction between two monomers on a chain that are close together in
space and have a non-zero relative velocity, and so on.
The simplest models for the internal viscosity force assume that it acts
only between neighbouring beads in a bead-spring chain, and depends on
the average relative velocities of these beads. Thus, for a bead $\mu$ that is
not at the chain ends,
\begin{equation}
{\bF}_{\mu}^{(iv)} = \varphi \, \biggl( {
({\br}_{\mu+1} - {\br}_{\mu})\,({\br}_{\mu+1} - {\br}_{\mu})
\over
\big\vert \, {\br}_{\mu+1} - {\br}_{\mu} \, \big\vert^2 }  \biggr)
\cdot \lmav \, { {\dot \br}}_{\mu+1} - { {\dot \br}}_{\mu} \, \rmav 
- \varphi \, \biggl( {
({\br}_{\mu} - {\br}_{\mu-1})\,({\br}_{\mu} - {\br}_{\mu-1})
\over
\big\vert \, {\br}_{\mu} - {\br}_{\mu-1} \, \big\vert^2 }  \biggr)
\cdot \lmav \, { {\dot \br}}_{\mu} - { {\dot \br}}_{\mu-1} \, \rmav 
\end{equation}
where $\varphi$ is the internal viscosity coefficient. A scaling theory 
for a more general model that accounts for internal friction between 
arbitrary pairs of monomers has also been developed~\cite{rabott}. 

The equation of motion for bead $\nu$ can consequently be written as,
\begin{equation}
- \zeta \, \Bigl[ \, \lmav { {\dot \br}}_{\nu} \rmav - \bv_0
 - \bk \cdot {\br}_{\nu} +  
\sum_{\mu} \, \bO_{\nu \mu} \cdot {\bF}_{\mu}^{(h)} \, \Bigr]
- k_{\rm B} T \, {\partial \, \ln \Psi \over
\partial {\br}_{\nu}} + {\bF}_{\nu}^{(\phi)}
+ {\bF}_{\nu}^{(iv)} = {\bzero}
\label{motion}
\end{equation}
Since ${\bF}_{\mu}^{(h)} = k_{\rm B} T \, ( {\partial \, \ln \Psi /
\partial {\br}_{\mu}} ) - {\bF}_{\mu}^{(\phi)}
- {\bF}_{\mu}^{(iv)}$, equation~(\ref{motion}) can be rearranged to give,
\begin{equation}
\lmav { {\dot \br}}_{\nu} \, \rmav = \bv_0 + \bk \cdot {\br}_{\nu} 
+ {1 \over \zeta} \, \sum_{\mu} \, \bgam_{\mu \nu} \cdot 
\biggl( \, - k_{\rm B} T \, {\partial \, \ln \Psi \over
\partial {\br}_{\mu}} + {\bF}_{\mu}^{(\phi)} 
+ {\bF}_{\mu}^{(iv)} \, \biggr) 
\label{eqrnudot}
\end{equation}
where $\bgam_{\mu \nu}$ is the dimensionless
{\it diffusion tensor}~\cite{birdb}, 
\begin{equation}
\bgam_{\mu \nu} = \delta_{\mu \nu } \, \bu + \zeta \, \bO_{\nu \mu} 
\end{equation}
By manipulating equation~(\ref{eqrnudot}), 
it is possible to rewrite the equation of motion in terms of the
velocities of the center of mass ${\br}_c$ and the
bead-connector vectors ${\bQ}_k$,
\begin{equation}
\lmav { {\dot \br}}_{c} \, \rmav = \bv_0 + \bk \cdot {\br}_{c} 
-  {1 \over N \zeta} \, \sum_{\nu,\mu,k} \, \overline B_{k \mu}
\, \bgam_{\mu \nu} \cdot 
\biggl( \, k_{\rm B} T \, {\partial \, \ln \Psi \over \partial {\bQ}_{k}}
+ {\partial \phi \over \partial {\bQ}_k} + {\bff}_{k}^{(iv)} \, \biggr) 
\label{eqrcdot}
\end{equation}
\begin{equation}
\lmav {\dot \bQ}_{j} \, \rmav = \bk \cdot {\bQ}_{j} 
- {1 \over \zeta} \, \sum_k \, \bA_{jk} \cdot 
\biggl( \, k_{\rm B} T \, {\partial \, \ln \Psi \over \partial {\bQ}_{k}}
+ {\partial \phi \over \partial {\bQ}_k} + {\bff}_{k}^{(iv)} \, \biggr) 
\label{eqqdot}
\end{equation}
where, ${\overline B}_{k \nu}$ is defined by, 
${\overline B_{k \nu}} = \delta_{k+1, \nu} - \delta_{k \nu}, $
the internal viscosity force, ${\bff}_{k}^{(iv)}$,  
in the direction of the connector vector ${\bQ}_k$ is, 
\begin{equation}
{\bff}_{k}^{(iv)} = \varphi \; { {\bQ}_k {\bQ}_k  \over
\big\vert \, {\bQ}_{k} \, \big\vert^2 }
\cdot \lmav \, {\dot \bQ}_{k} \, \rmav
\label{ivf}
\end{equation}
and the tensor $\bA_{jk}$ which accounts for the presence of
hydrodynamic interaction is defined by,
\begin{equation}
\bA_{jk} = \sum_{\nu, \, \mu} \, \overline B_{j \nu}
\, \bgam_{\mu \nu} \overline B_{k \mu}
= A_{jk} \bu + \zeta \bigl( \bO_{j,k} + \bO_{j+1,k+1}
- \bO_{j,k+1} - \bO_{j+1,k} \bigr) 
\end{equation}
Here, $A_{jk}$ is the Rouse matrix,
\begin{equation}
A_{jk}=\cases{ 2&for $\vert {j-k} \vert = 0 $,\cr
\noalign{\vskip3pt}
-1& for $\vert {j-k} \vert =1 $,\cr
\noalign{\vskip3pt}
0 & otherwise \cr} 
\end{equation}

In order to obtain the {\it diffusion} equation for a dilute solution of
bead-spring chains, the equation of motion derived here must be combined
with an equation of continuity.

\subsubsection{The diffusion equation} 
The equation of continuity or `probability conservation', which states
that a bead-spring chain that disappears from one configuration must
appear in another, has the form~\cite{birdb}, 
\begin{equation}
{\partial \, \Psi \over \partial t} = -\sum_{\nu} \,
{\partial \over \partial {\br}_{\nu}}
\cdot \lmav { {\dot \br}}_{\nu} \rmav \, \Psi
\end{equation}
The independence of $\Psi$ from the location of the center of mass for
homogeneous flows,
and the result ${\rm tr} \, \bk = 0$,  for an incompressible fluid,
can be shown to imply that the equation of continuity can be written in
terms of internal coordinates alone as~\cite{birdb},
\begin{equation}
{\partial \, \psi \over \partial t} = -\sum_j \,
{\partial \over \partial {\bQ}_j}
\cdot \lmav {\dot \bQ}_j \rmav \, \psi
\label{cont}
\end{equation}

The general diffusion equation which governs the time evolution
of the instantaneous configurational distribution function $\psi$,
in the presence of hydrodynamic interaction,
arbitrary spring and excluded volume forces, and an internal viscosity
force given by equation~(\ref{ivf}), is obtained by 
substituting the equation of motion for $\lmav {\dot \bQ}_j
\rmav$ from equation~(\ref{eqqdot}) into equation~(\ref{cont}). It has the
form, 
\begin{equation}
{\partial \, \psi \over \partial t} = - \sum_j \,
{\partial   \over \partial {\bQ}_j} \cdot \biggl( 
\bk \cdot {\bQ}_{j} 
- {1 \over \zeta} \, \sum_k \, \bA_{jk} \cdot 
\, \Bigl[ \, {\partial \phi \over \partial {\bQ}_k}
+ {\bff}_{k}^{(iv)} \, \Bigr]  \, \biggr) \, \psi 
+ {k_{\rm B} T \over \zeta} \, \sum_{j, \, k} \, 
{\partial \over \partial {\bQ}_j} \cdot \bA_{jk} 
\cdot {\partial \psi \over \partial {\bQ}_k}
\label{diff}
\end{equation}

Equations such as~(\ref{diff}) are also referred to as
{\em Fokker-Planck} or {\em Smoluchowski} equations in the literature.
The diffusion equation~(\ref{diff}) is the most fundamental equation of the
kinetic theory of dilute polymer solutions since a knowledge of $\psi$, 
for a flow field specified by $\bk$, would make it possible to evaluate
averages of various configuration dependent quantities and thereby permit
comparison of theoretical predictions with experimental observations.

The diffusion equation can be used to derive the time evolution equation
of the average of any arbitrary configuration dependent quantity,
$X (\,{\bQ}_1, \, \ldots, \, {\bQ}_{N-1} \,), $
by multiplying the left and right hand sides
of equation~(\ref{diff}) by $X$ and integrating both sides over all
possible configurations,
\begin{equation}
{d \, \avel X \aver \over dt} =
\sum_j \, \avel \, \bk \cdot {\bQ}_{j}
\cdot {\partial \, X \over \partial {\bQ}_{j}} \, \aver 
- { k_{\rm B} T \over \zeta} \, \sum_{j, \, k} \, \avel \, \bA_{jk} \cdot 
{\partial \, \ln \psi \over \partial {\bQ}_{k}}
\cdot {\partial \, X \over \partial {\bQ}_{j}} \, \aver 
- { 1 \over \zeta} \, \sum_{j, \, k} \, \avel \, \bA_{jk} \cdot 
\Bigl[ \, {\partial \phi \over \partial {\bQ}_k} + 
{\bff}_{k}^{(iv)} \, \Bigr]
\cdot {\partial \, X \over \partial {\bQ}_{j}} \, \aver
\label{avgx}
\end{equation}

Except for a situation where nearly all the important
microscopic phenomena are neglected, the diffusion equation~(\ref{diff})
is unfortunately in general analytically
insoluble. There have been very few attempts to directly solve  
diffusion equations with the help of a numerical solution
procedure~\cite{fan1,fan2}.
In this context it is worth bearing in mind that what are usually required are
averages of configuration dependent quantities. However,
in general even averages cannot be obtained exactly by solving
equation~(\ref{avgx}). As a result, it is common in most molecular
theories to obtain the averages by means of various approximations.

In order to examine the validity of
these approximations it is vitally important to compare the approximate
predictions of transport properties with the exact predictions of the models.
One of the ways by which exact numerical results may be
obtained is by adopting a numerical procedure based on the mathematical
equivalence of diffusion equations in polymer configuration space and
stochastic differential equations for the polymer configuration~\cite{ottbook}.
Instead of numerically solving the analytically intractable difusion equation for
the distribution function, stochastic trajectories can be generated
by {\it Brownian dynamics simulations} based on a numerical integration of the
appropriate stochastic differential equation. Averages calculated
from stochastic trajectories
(obtained as a solution of the stochastic differential equations), are
identical to the averages calculated from distribution functions 
(obtained as a solution of the diffusion equations).
It has now become fairly common for any new approximate molecular theory of a
microscopic phenomenon to establish the accuracy of the results with the
help of Brownian dynamics simulations. In this chapter, while results of
such simulations are cited, details of the development of
the appropriate stochastic differential equations are not discussed. A
comprehensive introduction to the development of stochastic differential
equations which are equivalent to given diffusion equations
for the probability density in configuration space, can be found in
the treatise by {\"O}ttinger~\cite{ottbook}. 

\subsubsection{The stress tensor}
The expression for the stress tensor in a dilute polymer solution was
originally obtained by the use of simple physical arguments 
which considered the various mechanisms that contributed to 
the flux of momentum across an 
oriented surface in the fluid~\cite{birdb}. The major mechanisms
considered were the transport of momentum by beads crossing the surface,
and the tension in the springs that straddle the surface. 
These physical arguments help to provide an intuitive understanding of
the origin of the different terms in the stress tensor expression. On the
other hand, such arguments are difficult to pursue in the presence of
complicated microscopic phenomena, and there is uncertainity about the
completeness of the final expression. An alternative approach to the 
derivation of the expression for the stress tensor has been to use
more fundamental arguments that consider the complete phase space 
of the polymeric fluid~\cite{birdb,cb}.

A very general expression for the polymer contribution to the stress tensor,
derived by adopting the complete phase space approach,
for models without constraints such as the bead-spring chain model, 
in the presence of hydrodynamic interaction and an arbitrary intramolecular
potential force, is the {\it modified Kramers} expression~\cite{birdb},
\begin{equation}
\btau^p = n_p \, \sum_{\nu} \, \avel \, ({\br}_{\nu} - {\br}_{c}) \, 
{\bF}_{\nu}^{(\phi)}\, \aver + (N-1) \,  n_p  k_{\rm B} T \, \bu 
\label{modkram}
\end{equation}
When rewritten in terms of the internal coordinates of a
bead-spring chain, equation~(\ref{modkram}) assumes a form called
the {\it Kramers} expression, 
\begin{equation}
\btau^p = - n_p \, \sum_j \, \avel \, {\bQ}_j \, 
{\partial \phi \over \partial {\bQ}_j} \, \aver
+  (N-1) \,  n_p  k_{\rm B} T \, \bu 
\label{kram}
\end{equation}

It is important to note that the presence of internal viscosity
has not been taken into account in the phase space theories 
used to derive the modified Kramers expression~(\ref{modkram}).
When examined from the standpoint of thermodynamic 
considerations, the proper form of the stress tensor in the 
presence of internal viscosity appears to be the Giesekus expression 
rather than the Kramers expression~\cite{schiebottiv}. Since 
predictions of models with internal viscosity are not 
considered in this chapter, the Giesekus expression is not 
discussed here. 

In order to evaluate the stress tensor, for various 
choices of the potential energy $\phi$, it turns out that 
it is usually necessary to 
evaluate the second moments of the bead connector vectors,
$\avel \, {\bQ}_j  {\bQ}_k \, \aver$. An equation that governs the
time evolution of the second moments can be obtained with the help
of equation~(\ref{avgx}). It has the form, 
\begin{eqnarray}
{d \over dt} \, \avel \, {\bQ}_j  {\bQ}_k \, \aver &=&
\bk \cdot \avel \, {\bQ}_j  {\bQ}_k \, \aver + 
\avel \, {\bQ}_j  {\bQ}_k \, \aver \cdot \bk^\dagger + 
{ 2 k_{\rm B} T \over \zeta} \, \avel \, \bA_{jk} \, \aver 
\nonumber \\ 
&-& { 1 \over \zeta} \, \sum_m \, \biggl\{ 
\avel \, {\bQ}_j \,  
\Bigl[ \, {\partial \phi \over \partial {\bQ}_m}
 +  {\bff}_{m}^{(iv)} \, \Bigr] \cdot \bA_{mk} \, \aver 
+ \avel \, \bA_{jm} \cdot 
\Bigl[ \, {\partial \phi \over \partial {\bQ}_m} + 
{\bff}_{m}^{(iv)} \, \Bigr] \cdot  {\bQ}_k \, \aver \; \biggr\} \hfill
\label{secmom}
\end{eqnarray}
The second moment equation~(\ref{secmom}), which is an ordinary
differential equation, is in general not a closed
equation for $\avel \, {\bQ}_j  {\bQ}_k \, \aver$, 
since it invoves higher order moments on
the right hand side. 

Within the context of the molecular theory developed thus far, it is
clear that the prediction of the rheological properties of dilute
polymer solutions with a bead-spring chain model usually 
requires the solution
of the second moment equation~(\ref{secmom}). To date however,
there are no solutions to the general second moment equation~(\ref{secmom})
which simultaneously incorporates the microscopic phenomena of
hydrodynamic interaction, excluded volume, non-linear spring forces and
internal viscosity. Attempts have so far been restricted to treating a
smaller set of combinations of these phenomenon. The simplest 
molecular theory, based on a bead-spring chain
model, for the prediction of the
transport properties of dilute polymer solutions is the Rouse model. The
Rouse model neglects all the microscopic phenomenon listed above, and
consequently fails to predict many of the observed features of dilute
solution behavior. In a certain sense, however, it provides the framework
and motivation for all further improvements in the molecular theory.
The Rouse model and its predictions are introduced below, 
while improvements in the treatment of hydrodynamic interactions 
alone are discussed subsequently.   

\subsection{The Rouse model}
The Rouse model assumes that the springs of the bead-spring chain are
governed by a Hookean spring force law. The only solvent-polymer
interactions treated are that of hydrodynamic drag and Brownian bombardment.
The diffusion equation~(\ref{diff}) with the effects of 
hydrodynamic interaction, excluded volume and internal viscosity 
neglected, and with a Hookean spring force law, has the form, 
\begin{equation}
{\partial \, \psi \over \partial t} = - \sum_j \,
{\partial   \over \partial {\bQ}_j} \cdot \biggl( 
\bk \cdot {\bQ}_{j} 
- {H \over \zeta} \, \sum_k \, A_{jk} \, {\bQ}_k
\, \biggr) \, \psi
+{k_{\rm B} T \over \zeta} \, \sum_{j, \, k} \, A_{jk} \; 
{\partial \over \partial {\bQ}_j} \cdot
{\partial \psi \over \partial {\bQ}_k}
\label{rousdiff}
\end{equation}

The diffusion equation~(\ref{rousdiff}) has an analytical solution
since it is linear in the bead-connector vectors. It is satisfied by a
Gaussian distribution, 
\begin{equation}
\psi \, ({\bQ}_1, \ldots, {\bQ}_{N-1}) \, = \, {\cal N} (t) \,
\exp \big[ - {1 \over 2} \sum_{j, \, k} \, {\bQ}_j \cdot 
({\bs}^{- 1})_{jk}
\cdot {\bQ}_k \big]
\label{gauss}
\end{equation}
where ${\cal N}(t)$ is the normalisation factor, and the tensor $\bsjk$
which uniquely characterises the Gaussian distribution is identical to the
second moment,
\begin{equation}
\bsjk \equiv \avel {\bQ}_j{\bQ}_k \aver 
\end{equation}
Note that the tensors $\bsjk$ are not symmetric, but satisfy the
relation $ \bsjk=\bs_{kj}^T $. 
(Further information on linear diffusion equations and Gaussian
distributions can be obtained in the extended discussion in Appendix A
of~\cite{ottca2}). 

Since the intramolecular potential in the Rouse model is only due to the
presence of Hookean springs, it is straight forward to see that
the Kramers expression for the stress tensor $\btau^p$, is given by,
\begin{equation}
\btau^p=- n_p H \, \sj \, \bs_{jj} + (N-1) \, n_p  k_{\rm B} T \, \bu 
\label{rkram}
\end{equation}
The tensors $\bs_{jj}$ are obtained by solving the
second moment equation~(\ref{secmom}), which becomes a closed equation
for the second moments when the Rouse assumptions are made.
It has the form, 
\begin{equation}
{d \over dt} \,  \bsjk - \bk \cdot  \bsjk 
- \bsjk \cdot \bk^\dagger  =  
{ 2 k_{\rm B} T \over \zeta} \,  A_{jk} \, \bu  
- { H \over \zeta} \, \sum_m \, \bigl[ \;  \bs_{jm} \, A_{mk}
+  A_{jm} \, \bs_{mk} \; \bigr]
\label{rsm}
\end{equation}
Note that the solution of
equation~(\ref{rsm}) also leads to the complete specification of
the Gaussian configurational distribution function $\psi$.

A {\it Hookean dumbbell } model, which is the simplest example of a bead-spring
chain, is obtained by setting $N=2$. It is often used for preliminary
calculations since its simplicity makes it possible to obtain analytical
solutions where numerical solutions are unavoidable for longer chains.
For such a model, substituting for $\bs_{11}$ in terms of $\btau^p$ from
equation~(\ref{rkram}) into equation~(\ref{rsm}), leads to following
equation for the polymer contribution to the stress tensor,
\begin{equation}
\btau^p + \lambda_H \, \btau^p_{(1)}
=  -  n_p k_{\rm B} T \lambda_H \, {\dot \bgam}
\label{hooktau}
\end{equation}
where $\btau^p_{(1)}  =  {d \, \btau^p / dt}  -
\bk \cdot  \btau^p - \btau^p \cdot \bk^\dagger, $ is the
convected time derivative~\cite{birdb} of $\btau^p$, and
$\lambda_H= (\zeta /4 H)$ is a time constant. Equation~(\ref{hooktau})
indicates that a Hookean dumbbell model with the Rouse assumptions
leads to a convected Jeffreys model or Oldroyd-B model as
the constitutive equation for a dilute polymer solution. This is perhaps
the simplest coarse-grained microscopic model capable of reproducing
some of the macroscopic rheological properties of dilute polymer
solutions. 

In the case of a bead-spring chain with $N > 2$, 
it is possible to obtain a similar
insight into the nature of the stress tensor by introducing {\it normal
coordinates}. These coordinates help to decouple the connector vectors
${\bQ}_1, \ldots,{\bQ}_{N-1}$, which are coupled to each other
because of the Rouse matrix. 

The connector vectors are mapped to a new set of normal
coordinates, ${\bQ}_1^\prime,\ldots,{\bQ}_{N-1}^\prime$
with the transformation,
\begin{equation}
{\bQ}_j=\sk \,\pijk {\bQ}_k^\prime
\label{normap}
\end{equation}
where, $\pijk$ are the elements of an orthogonal matrix
with the property
\begin{equation}
(\Pi^{-1})_{jk}=\Pi_{kj}, \, \, {\rm such \; that,} \, \, 
\sm \, \pimj\pimk= \delta_{jk}
\end{equation}

The orthogonal matrix $\pijk$, which will henceforth be referred to as
the Rouse orthogonal matrix, diagonalises the Rouse matrix $A_{jk}$,
\begin{equation}
\sum_{j, \, k} \, \Pi_{ji}  A_{jk} \Pi_{kl} = a_l \delta_{il}
\label{roueig}
\end{equation}
where, the Rouse eigenvalues $a_l$ are given by $a_l = 4 \sin^2 (l \pi /2 N) $. 
The elements of the Rouse orthogonal matrix are given by the expression,
\begin{equation}
\pijk=\sqrt{2 \over N} \, \sin
\Biggl({jk \pi \over N } \Biggr) 
\end{equation}

The diffusion equation in terms of these normal coordinates, admits a
solution for the configurational distribution function of the form~\cite{birdb} 
\begin{equation}
\psi \, ({\bQ}_1^\prime,\ldots,{\bQ}_{N-1}^\prime)= \prod_{k=1}^{N-1}
\, \psi_k \, ({\bQ}_k^\prime) 
\end{equation}
As a consequence, the diffusion equation becomes uncoupled and can be
simplified to $(N-1)$ diffusion equations, one for each of
the $\psi_k \, ({\bQ}_k^\prime)$. Since the
${\bQ}_k^\prime$ are independent of each other, all the covariances 
$\avel {\bQ}_j^\prime {\bQ}_k^\prime \aver$ with $j \ne k$ are zero,
and only the $(N-1)$ variances $\bs^\prime_{j} \equiv
\avel {\bQ}_j^\prime {\bQ}_j^\prime \aver$ are non-zero. Evolution equations 
for the variances $\bs^\prime_{j} $ can then be derived from these
uncoupled diffusion equations with the help of a procedure
similar to that used for the derivation
of equation~(\ref{secmom}).

The stress tensor is given in terms of $\bs^\prime_{j} $ by the expression,
\begin{equation}
\btau^p= \sj \, \btau_j^p
\label{tausum}
\end{equation}
where,
\begin{equation}
\btau_j^p=- n_p H \, \bs_j^\prime + n_p  k_{\rm B} T \, \bu
\label{tauj}
\end{equation}
On substituting for $\bs_j^\prime$ in terms of $\btau_j^p$
from equation~(\ref{tauj}) into the evolution equation for $\bs_j^\prime$,
one obtains,
\begin{equation}
\btau_j^p + \lambda_j \, \btau^p_{j \, (1)}
=  -  n_p k_{\rm B} T \lambda_j \, {\dot \bgam}
\label{rautau}
\end{equation}
where, the relaxation times $\lambda_j$ are given by
$\lambda_j = (\zeta /2 H \, a_j)$. Consequently,  
each of the $\btau_j^p$ satisfy an equation identical to
equation~(\ref{hooktau}) for the polymer contribution to the stress tensor in
a Hookean dumbbell model. The Rouse model, therefore, leads to a constitutive
equation that is a multimode generalization of the convected Jeffreys or
Oldroyd B model.

It is clear from above discussion that the process of
transforming to normal coordinates enables one to derive a closed form
expression for the stress tensor, and to  
gain the insight that the Rouse chain with $N$ beads has
$N$ independent relaxation times which describe the different relaxation
processes in the chain, from the entire chain to successively smaller
sub-chains. It is straight forward to show that for large $N$,
the longest relaxation times $\lambda_j$ scale with chain length as $N^2$.

A few important transport property predictions which 
show the limitations of the Rouse model
are considered briefly below. It is worth noting that since the Rouse
model does not include the effect of excluded volume, its predictions
are restricted to dilute solutions of polymers in theta solvents. This
restriction is infact applicable to all the models of hydrodynamic interaction 
treated here. 

In steady simple shear flow, with $\bk(t)$ given by
equation~(\ref{ssf1}), the three independent material
functions that characterise such flows are~\cite{birdb},
\begin{equation}
\eta_{p} =  n_p  k_{\rm B} T \, \sj \lambda_j \, ; \quad  
\Psi_{1} = 2 n_p  k_{\rm B} T \, \sj \lambda^2_j \, ; \quad 
\Psi_{2} = 0 
\label{etap}
\end{equation}
It is clear that the Rouse model accounts for the presence of viscoelasticity
through the prediction of a nonzero first normal stress difference in
simple shear flow. However, it does not predict the nonvanishing of
the second normal stress difference, and the shear rate dependence
of the viscometric functions.

From the definition of intrinsic viscosity~(\ref{invis}) and the fact
that $\rho_p \sim N \, n_p$, it follows from equation~(\ref{etap})
that for the Rouse model,
$\lbrack \eta \rbrack_0 \sim  N$. This is at variance with the
experimental results discussed earlier, and displayed in equation~(\ref{sc1}). 
It is also straight forward to see that the Rouse model predicts that the
characteristic relaxation time scales as the square of the chain length, 
$\lambda_p \sim N^2$.

In small amplitude oscillatory shear, $\bk(t)$ is given by
equation~(\ref{usf3}), and expressions for the material functions
$G^\prime$ and $G^{\prime \prime}$ in terms of the relaxation times
$\lambda_j$ can be easily obtained~\cite{birdb}. In the intermediate
frequency range, where as discussed earlier, experimental results
indicate that both $G^{\prime}$ and $G^{\prime\prime}$
scale as $\omega^{2/3}$, the Rouse model predicts a scaling 
$\omega^{1/2}$~\cite{larson}.

The translational diffusion coefficient $D$ for a bead-spring chain
at equilibrium can be obtained by finding the average friction
coefficient $Z$ for the entire chain in a quiescent solution, 
and subsequently using the Nernst-Einstein equation,
$D=k_{\rm B} T \, Z^{-1}$~\cite{birdb}. It can be shown that
for the Rouse model $Z = \zeta \, N$, {\it ie.} the
total friction coefficient of the chain is a sum of the individual bead friction
coefficients. As a result, the Rouse model predicts that the diffusion
coefficient scales as the inverse of the molecular weight. This is not
observed in dilute solutions. Instead experiments indicate the scaling
depicted in equation~(\ref{sc2}).

The serious shortcomings of the Rouse model highlighted above have been
the motivation for the development of more refined molecular theories. 
The scope of this chapter is restricted to reviewing recent advances 
in the treatment of hydrodynamic interaction. 

\section{HYDRODYNAMIC INTERACTION}
Hydrodynamic interaction, as pointed out earlier, is a long range
interaction between the beads
which arises because of the solvent's capacity to propagate one bead's
motion to another through perturbations in
its velocity field. It was first introduced into framework of polymer
kinetic theory by Kirkwood and Riseman~\cite{kirkrise}. As we have seen in
the development of the general diffusion equation above, it is reasonably
straight forward to include hydrodynamic interaction into the framework of the
molecular theory. However, it renders the resultant equations analytically
intractable and as a result, various attempts have been made in order to solve
them approximately. 

In this section, we review the various approximation schemes introduced
over the years. The primary test of an approximate model is of
course its capacity to
predict experimental observations. The accuracy of the approximation
however, can only be assessed by checking the proximity of the
approximate results to the exact numerical results obtained by
Brownian dynamics simulations. Finally, the usefulness of an approximation
depends on its computational intensity. The individual features and
deficiencies of the different approximations will be examined in the
light of these observations. 

In the presence of hydrodynamic interaction, and with excluded volume
and internal viscosity neglected, a bead-spring chain with Hookean
springs has a configurational distribution function $\psi$ that must
satisfy the following simplified form of the diffusion equation~(\ref{diff}),
\begin{equation}
{\partial \psi \over \partial t} = - \sj {\partial \over \partial {\bQ}_j}
\cdot \biggl( \bk \cdot {\bQ}_j - {H \over \zeta}
\, \sk \bA_{jk} \cdot {\bQ}_k \biggr) \psi  
+ {k_B T \over \zeta} \, \sum_{j, \, k} 
{\partial \over \partial {\bQ}_j} \cdot \bA_{jk} \cdot 
{\partial \psi \over \partial {\bQ}_k} 
\label{hidiff}
\end{equation}
while the second moment equation~(\ref{secmom}) assumes the form,
\begin{equation}
{d \over dt} \avel {\bQ}_j{\bQ}_k \aver =
\bk \cdot \avel {\bQ}_j{\bQ}_k \aver+\avel {\bQ}_j
{\bQ}_k \aver \cdot \bk^\dagger
+ {2 k_B T \over \zeta}\, \avel\bA_{jk} \aver 
- {H \over \zeta} \, \sm \biggl\lbrack \,\avel {\bQ}_j{\bQ}_m \cdot
\bA_{mk} \aver + \avel \bA_{jm} \cdot {\bQ}_m{\bQ}_k \aver \,
\biggr\rbrack
\label{hisecmom}
\end{equation}
Equation~(\ref{hisecmom}) is not a
closed equation for the second moments since it involves more complicated
moments on the right hand side. This is the central problem of
all molecular theories which attempt to predict the rheological properties of
dilute polymer solutions and that incorporate hydrodynamic
interaction. The different approximate treatments of hydrodynamic interaction,
which are discussed roughly chronologically below,
basically reduce to finding a suitable
closure approximation for the second moment equation.

\subsection{ The Zimm model} 
The Zimm model was the first attempt at improving the Rouse model by
introducing the effect of hydrodynamic interaction in a 
preaveraged or equilibrium-averaged form. 
The preaveraging approximation has been very frequently used
in polymer literature since its introduction by Kirkwood and
Riseman~\cite{kirkrise}. The approximation consists of evaluating the 
average of the hydrodynamic tensor with the equilibrium
distribution function~(\ref{equidis}), and 
replacing the hydrodynamic interaction tensor $ \bA_{jk}$, wherever
it occurs in the governing equations,  
with its equilibrium average ${\widetilde A}_{jk}$.
(Note that the incorporation of the effect of hydrodynamic interaction
does not alter the equilibrium distribution function, which is still given
by~(\ref{equidis}) for bead-spring chains with Hookean springs.)
The matrix ${\widetilde A}_{jk}$ is called the modified Rouse matrix,
and is given by,
\begin{equation}
{\widetilde A}_{jk}=A_{jk}+\sqrt{2} \, h^*
\Biggl({2 \over \sqrt{\vert j-k \vert}}-{1 \over \sqrt{\vert j-k-1 \vert}}
-{1 \over \sqrt{\vert j-k+1 \vert}} \Biggr) 
\end{equation}
where, $h^*=a {\sqrt {(H / \pi k_B T)}}$  is the hydrodynamic interaction
parameter. The hydrodynamic interaction parameter is approximately equal 
to the ratio of the bead radius to the equilibrium root mean square 
length of a single spring of the bead-spring chain.
This implies that  $h^* < 0.5$, since the beads cannot overlap.
Typical values used for $h^*$ are in the range
$0.1 \le h^* \le 0.3$~\cite{ottca2}. 

By including the hydrodynamic interaction in an averaged form,
the diffusion equation remains linear in the connector vectors,
and consequently is satisfied by a Gaussian distribution~(\ref{gauss})
as in the Rouse case. However, the covariance tensors $\bsjk$ are
now governed by the set of differential equations~(\ref{rsm}) 
with the Rouse matrix $A_{jk}$ replaced with the modified Rouse
matrix ${\widetilde A}_{jk}$. Note that this modified second moment equation
is also a closed set of equations for the second moments. 

As in the Rouse case, it is possible to simplify the solution of the
Zimm model by carrying out a diagonalisation procedure. This is
achieved by mapping the connector vectors to normal coordinates, as
in~(\ref{normap}), but in this case the Zimm  orthogonal matrix
$\Pi_{jk}$, which diagonalises the modified Rouse matrix,
\begin{equation}
\sum_{j, \, k} \, \Pi_{ji} {\widetilde A}_{jk} \Pi_{kl} =
{\widetilde a}_l \delta_{il}
\label{zimeig}
\end{equation}
must be found numerically for $N > 4$. Here, ${\widetilde a}_l$ are
the so called Zimm eigenvalues. The result of this procedure is to
render the diffusion equation 
solvable by the method of separation of variables.
Thus, as in the Rouse case, only the $(N-1)$ transformed coordinate
variances $\bs^\prime_{j} $ are non-zero, and differential equations governing 
these variances can be derived by manipulating the uncoupled
diffusion equations.

The diagonalisation procedure enables the polymer contribution to
the stress tensor $\btau^p$ in the Zimm model to be expressed
as a sum of partial stresses $\btau_j^p$ as in equation~(\ref{tausum}),
but the $\btau_j^p$ now satisfy equation~(\ref{rautau}) 
with the `Rouse' relaxation times $\lambda_j$ replaced with `Zimm'
relaxation times ${\widetilde \lambda}_j$. The Zimm relaxation times
are defined by ${\widetilde \lambda}_j = (\zeta /2 H \, {\widetilde a}_j)$. 

From the discussion above, it is clear that the Zimm model differs from
the Rouse model only in the spectrum of relaxation times. As we shall
see shortly, this leads to a significant improvement in the prediction
of linear viscoelastic properties and the scaling of
transport properties with molecular weight in theta solvents. 
The Zimm model therefore establishes
unequivocally the importance of the microscopic phenomenon of
hydrodynamic interaction. On the other hand, it does not lead to any
improvement in the prediction of nonlinear properties, and consequently  
subsequent treatments of hydrodynamic interaction have concentrated
on improving this aspect of the Zimm model. 

By considering the long chain limit of the Zimm model,
{\it ie.,} $N \to \infty$, it is possible to discuss the universal
properties predicted by the model. The various power law dependences
of transport properties on molecular weight, characterised by universal
exponents, and universal ratios formed from the prefactors of these 
dependences can be obtained. These predictions are
ideal for comparison with experimental data on high molecular weight
polymer solutions since they are parameter free. We shall discuss some
universal exponents predicted by the Zimm model below, while universal
ratios are discussed later in the chapter. 

As mentioned above, the first noticeable change upon
the introduction of hydrodynamic interaction is the change
in the relaxation spectrum. In the long chain limit, the longest relaxation
times ${\widetilde \lambda}_j$ scale with 
chain length as $N^{3/2}$~\cite{ottbook},
whereas we had found earlier that the chain length dependence of the
longest relaxation times in the Rouse model was $N^2$.

In steady simple shear flow, the Zimm model like the Rouse model, fails
to predict the experimentally observed occurance of non-zero
second normal stress differences and the experimentally observed
shear rate dependence of the
viscometric functions. It does however lead to an improved prediction of the
scaling of the zero shear rate intrinsic viscosity with molecular
weight, $\lbrack \eta \rbrack_0 \sim  N^{1/2}$. This prediction is in
agreement with experimental results for the Mark-Houwink exponent 
in theta solvents (see equation~(\ref{sc1})). As with the longest
relaxation times, the characteristic relaxation time
$\lambda_p \sim N^{3/2}$.  

In small amplitude oscillatory shear, the Zimm model predicts  that
the material functions $G^\prime$ and $G^{\prime \prime}$
scale with frequency as $\omega^{2/3}$ in the intermediate
frequency range. This is in exceedingly good agreement with experimental
results~\cite{birdb,larson}.

The translational diffusion coefficient $D$ for chainlike molecules
at equilibrium, with preaveraged hydrodynamic interaction,
was originally obtained by Kirkwood~\cite{kirkrise}. Subsequently,
several workers obtained a correction to the Kirkwood diffusion
coefficient for the Zimm model~\cite{otttd}. The exact results differ 
by less than 2\% from the Kirkwood value for all values of the chain
length and $h^*$. Interestingly, three different approaches 
to obtaining the diffusion coefficient, namely, the Nernst-Einstein equation, 
the calculation of the mean-square displacement caused by Brownian forces, and 
the study of the time evolution of concentration gradients, lead to 
identical expressions for the diffusion coefficient~\cite{otttd}. In 
the limit of very long chains, it can be shown that $D \sim N^{- 1/2}$. 
The Zimm model therefore gives the correct dependence of translational 
diffusivity on molecular weight in theta solvents. 

The Zimm result for the translational diffusivity has been
traditionally interpreted to mean that the polymer coil in
a theta solvent behaves like a rigid
sphere, with radius equal to the root mean square end-to-end distance.
This follows from the fact that the diffusion coefficient for a
rigid sphere scales as the inverse of the radius of the sphere, and
in a theta solvent, 
$\avel \, r^2 \, \aver_{\rm eq}$ scales with chain length as $N$.
The solvent inside the coil is believed to be dragged along with the
coil, and the inner most beads of the bead-spring chain
are considered to be shielded from the velocity field due to the presence of 
hydrodynamic interaction~\cite{yamakawa,larson}.
This intuitive notion has been used to point out the difference between
the Zimm and the Rouse model, where all the $N$ beads of the polymer
chain are considered to be exposed to the applied velocity field.
Recently, by explicitly calculating the velocity field inside a
polymer coil in the Zimm model, \Ott~\cite{ottvel} has shown 
that the solvent motion inside a polymer coil is different
from that of a rigid sphere throughout the polymer coil, and that
shielding from the velocity field occurs only to a certain extent. 

\subsection{ The consistent averaging approximation} 
The first predictions of shear thinning were obtained when
hydrodynamic interaction was treated in a more precise manner than 
that of preaveraging the hydrodynamic interaction tensor. 
In order to make the diffusion equation~(\ref{hidiff}) linear in the
connector vectors, as pointed out earlier, 
it is necessary to average the hydrodynamic interaction 
tensor. However, it is not necessary to
preaverage the hydrodynamic interaction tensor with the equilibrium
distribution. On the other hand, the average can be carried out with 
the non-equilibrium distribution function~(\ref{gauss}). The linearity 
of the diffusion equation ensures that its solution is a 
Gaussian distribution. \Ott~\cite{ottca1,ottca2} suggested that the
hydrodynamic interaction tensor occuring in the diffusion equation 
be replaced with its non-equilibrium average. Since it is necessary to know
the averaged hydrodynamic interaction tensor in order to find the
non-equilibrium distribution function, both the averaged
hydrodynamic interaction tensor and the non-equilibrium distribution
function must be obtained in a {\it self-consistent} manner.

Several years ago, Fixman~\cite{fixman} introduced an iterative 
scheme (beginning with the equilibrium distribution function), 
for refining the distribution function with which to carry out 
the average of the hydrodynamic interaction tensor. The
self-consistent scheme of \Ott is recovered if the iterative procedure 
is repeated an infinite number of times. However, Fixman carried out the
iteration only upto one order higher than the preaveraging stage. 

The average of the hydrodynamic interaction tensor evaluated with
the Gaussian distribution~(\ref{gauss})
is an $(N-1) \times (N-1)$ matrix with tensor components, $\bAbjk$, defined by,
\begin{equation}
\bAbjk = A_{jk}\,\bu  +
\sqrt{2} h^* \, \Biggl\lbrack
\, {\bH(\bsh_{j,k})\over\sqrt{\vert j-k \vert}} 
+ {\bH(\bsh_{j+1,k+1})\over\sqrt{\vert j-k \vert}} 
- {\bH(\bsh_{j,k+1})\over\sqrt{\vert j-k-1 \vert}} 
- {\bH(\bsh_{j+1,k})\over\sqrt{\vert j-k+1 \vert}}
\, \Biggr\rbrack  
\label{Abar}
\end{equation}
where the tensors $\bshmn$ are given by,
\begin{equation}
\bshmn= \bshmn^\dagger = \bsh_{\nu \mu} = {1 \over \vert \mu - \nu \vert }\,
{H \over k_B T }
\,\, \sum_{j,k \,= \,\min(\mu,\nu)}^{ \max(\mu,\nu)-1} \, \bsjk
\end{equation}
and the function of the second moments, ${\bH }(\bs)$ is,
\begin{equation}
\bH(\bs)={3 \over 2 (2 \pi)^{3/2}} \int d {\bfk} \,{1 \over k^2}\,
\biggl( \bu - {{\bfk}{\bfk} \over k^2} \biggr) \,\exp(- {1 \over 2} {\bfk}
\cdot \bs \cdot {\bfk})
\label{hfunc}
\end{equation}
Note that the convention $\bH(\bsh_{jj})/0=0$ has been adopted in
equation~(\ref{Abar}) above. 

The self-consistent closure approximation therefore consists of replacing
the hydrodynamic interaction tensor $ \bA_{jk}$ in
equation~(\ref{hisecmom}) with its non-equilibrium
average $\bAbjk$. As in the earlier approximations, this leads to
a system of  $(N-1)^2$
coupled ordinary differential equations for the components of the covariance
matrix $\bsjk$. Their solution permits the evaluation of the
stress tensor through the Kramers expression~(\ref{rkram}), and as a
consequence all the relevant material functions.

Viscometric functions in steady simple shear flows
were obtained by  \Ott~\cite{ottca1} for chains with $N \leq 25$ beads, 
while material functions in start-up of steady shear flow, cessation 
of steady shear flow, and stress relaxation after step-strain were 
obtained by Wedgewood and \Ott~\cite{wedgeott} for chains with 
$N \leq 15$ beads. The latter authors also include consistently-averaged 
FENE springs in their model. 

Shear rate dependent viscometric functions, and a nonzero 
{\it positive} second normal stress difference are predicted by the 
self-consistent averaging approximation; a marked improvement over 
the predictions of the Zimm model. Both the reduced 
viscosity and the reduced first normal stress difference initially 
decrease with increasing shear rate. However, for long enough chains, 
they begin to rise monotonically at higher values of the reduced shear 
rate $\beta$. This rise is a consequence of the weakening of
hydrodynamic interaction in strong flows due to an increase in 
the separation between the beads of the chain. 
With increasing shear rate, the material functions 
tend to the shear rate independent Rouse values, 
which (for long enough chains), are higher than the zero shear rate 
consistently-averaged values. The prediction of 
shear thickening behavior is not in agreement with the shear thinning that is  
commonly observed experimentally. However, as mentioned earlier, 
some experiments with very high molecular weight systems seem to suggest 
the existence of shear thinning followed by shear thickening followed 
again by shear thinning as the shear rate is increased. While only 
shear thickening at high shear rates is predicted with Hookean springs, the 
inclusion of consistently-averaged FENE springs in the model leads to 
predictions which are qualitatively in agreement with these observations, 
with the FENE force becoming responsible for the shear thinning at very 
high shear rates~\cite{wedgeott,kish}. 

The means of examining the accuracy of various approximate treatments of
hydrodynamic interaction was established when the problem was solved 
exactly with the help of Brownian dynamics simulations with full
hydrodynamic interaction included~\cite{zylottga,zylkaga}. These 
simulations reveal that while the
predictions of the shear rate dependence of the viscosity and
first normal stress difference by the self-consistent averaging procedure
are in qualitative agreement with the Brownian dynamics simulations, they
do not agree quantitatively. Further, in contrast to the 
consistent-averaging prediction, at low shear rates, a
negative value for the second normal stress difference is obtained. 
As noted earlier,  the sign of the second normal stress difference 
has not been conclusively established~\cite{birdott}. 

The computational intensity of the consistent-averaging approximation 
leads to an upper bound on the length of chain that can be examined. As a
result, it is not possible to discuss the universal shear rate 
dependence of the viscometric functions predicted by it. On the other hand,
it is possible to come to certain general conclusions regarding the 
nature of the stress tensor in the long chain limit, and to predict 
the zero shear rate limit of certain universal ratios~\cite{ottca1}. 
Thus, it is possible to show the important result that the polymer 
contribution to the stress tensor depends only on a 
length scale and a time scale, and not on the strength of the hydrodynamic 
interaction parameter $h^*$. In the long chain limit,   
$h^*$ can be absorbed into the basic time constant, and it does not occur 
in any of the non-dimensional ratios. Indeed this is also true of the 
finite extensibility parameter $b$, which can also be shown to have 
no influence on the long chain rheological properties~\cite{ottca2}. 
The long chain limit of the consistent-averaging 
approximation is therefore a parameter free model.  

It is possible to obtain an explicit representation of the modified 
Kramers matrix for infinitely long chains by introducing continuous 
variables in place of discrete indices~\cite{ottca1}. This enables the 
analytical calculation of various universal ratios predicted by the 
consistent-averaging approximation. These predictions are discussed 
later in this chapter. However, two results are worth highlighting here. 
Firstly, it can be shown explicitly that the leading order corrections 
to the large $N$ limit of the various universal ratios are of 
order $(1/\surd N)$, and secondly, there is a special value 
of $h^*=0.2424...,$ at which the leading order corrections are 
of order $(1/ N)$. These results have proven 
to be very useful for subsequent numerical exploration of the long 
chain limit in more accurate models of the hydrodynamic interaction. 

Short chains with consistently-averaged hydrodynamic interaction,
as noted earlier, do not show shear thickening behavior;  
this aspect is revealed only with increasing chain length. Furthermore,
it is not clear with the kind of chain lengths that can be 
examined, whether the minimum in the viscosity and first normal stress
curves continue to exist in the long chain limit~\cite{ottca1}.
The examination of long chain behavior is therefore important since
aspects of polymer solution behavior might be revealed that are   
otherwise hidden when only short chains are considered. 
The introduction of the {\it decoupling approximation} 
by Magda, Larson and Mackay~\cite{magda} and Kishbaugh and McHugh~\cite{kish} 
made the examination of the shear rate dependence of long chains feasible. The 
decoupling approximation retains the accuracy of the self-consistent
averaging procedure, but is much more computationally efficient.

\subsection{The decoupling approximation} 
The decoupling approximation introduced by Magda {\it et al.}~\cite{magda}
and Kishbaugh and McHugh~\cite{kish} (who use FENE springs in place of
the Hookean springs of Magda {\it et al.})
consists of extending the `diagonalise and decouple' procedure
of the Rouse and Zimm theories to the case of the self consistently
averaged theory. They first transform the connector vectors $\bQ_j$
to a new set of coordinates $\bQ_j^\prime$
using the time-invariant Rouse orthogonal matrix $\pijk$. 
(Kishbaugh and McHugh also use the Zimm orthogonal matrix). 
The same orthogonal matrix $\pijk$ is then assumed to 
diagonalise the matrix of {\it tensor} components $\bAbjk$.
While the process of diagonalisation was exact in the Rouse and Zimm theories, 
it is an approximation in the case of the decoupling approximation. 
It implies that even in the self consistently averaged theory the 
diffusion equation can be solved by the method of separation of variables, 
and only the $(N-1)$ transformed coordinate variances $\bs^\prime_{j} $ 
are non-zero. The differential equations governing these variances 
can be derived from the uncoupled diffusion equations and solved
numerically. The appropriate material functions are then obtained 
using the Kramers expression in terms of the transformed coordinates, namely,
equations~(\ref{tausum}) and~(\ref{tauj}). 
  
The decrease in the number of differential equations to be solved,
from $(N-1)^2$ for the covariances $\bsjk$ to $(N-1)$ for the variances 
$\bs^\prime_j$, is suggested by Kishbaugh and McHugh as the 
reason for the great reduction in computational time achieved by the
decoupling approximation. Prakash and {\" O}ttinger~\cite{prakott} 
discuss the reasons why this argument is incomplete, and point out 
the inconsistencies in the decoupling procedure. Furthermore, 
since the results are
only as accurate as the consistent averaging approximation, the decoupling 
approximation is not superior to the consistent averaging method. 
However, these papers are important 
since the means by which a reduction in computational intensity may 
be achieved, without any significant sacrifice in accuracy, 
was first proposed in them. Further, the persistence of the 
minimum in the viscosity and first normal stress curves even for 
very long chains, and the necessity of including
FENE springs in order to generate predictions in qualitative agreement 
with experimental observations in high molecular weight systems, is clearly 
elucidated in these papers. 

\subsection{The Gaussian approximation}
The closure problem for the second moment equation is solved in the
preaveraging assumption of Zimm, and in the self consistent averaging
method of {\" O}ttinger, by replacing the tensor $\bA_{jk}$ with an average.
As a result, fluctuations in the hydrodynamic interaction are neglected.
The Gaussian approximation~\cite{ottga,zylottga,wedgega,zylkaga}  
makes no assumption with regard to the hydrodynamic interaction,
but assumes that the solution of the diffusion equation~(\ref{hidiff}) 
may be approximated by a Gaussian distribution~(\ref{gauss}).
Since all the complicated averages on the right hand side of the 
second moment equation~(\ref{hisecmom}) can be reduced to functions
of the second moment with the help of the Gaussian distribution, 
this approximation makes it a closed equation for the second moments.

The evolution equation for the covariances $\bsjk$ is given by,
\begin{eqnarray}
{d \over dt} \bsjk &=& \bk \cdot \bsjk
+ \bsjk \cdot \bk^T 
+{2 k_B T \over \zeta}\,\, \bAbjk - {H \over \zeta} \, \sm \, \lbrack
\bs_{jm} \cdot \bAb_{mk} + \bAb_{jm} \cdot \bs_{mk} \rbrack
\nonumber \\
&-& {H \over \zeta}\,{H \over k_B T}\, \sum_{m, \, l, \, p} \, \lbrack
\bs_{jl} \cdot \bG_{lp,mk} : \bs_{pm} + \bs_{mp} : \bG_{lp,jm} \cdot
\bs_{lk} \rbrack
\label{gasecmom}
\end{eqnarray}
where, the $(N-1)^2 \times (N-1)^2$ matrix with fourth rank
tensor components, $\bG_{lp,jk}$, is defined by,
\begin{eqnarray}
\bG_{lp,jk}&=&{3 \sqrt{2} \, h^* \over 4} \,\Biggl\lbrack \,
{\theta(j,l,p,k)\, \bK(\bsh_{j,k})+\theta(j+1,l,p,k+1)\, \bK(\bsh_{j+1,k+1})
\over \sqrt{{\vert j-k \vert}^3}} \nonumber \\
&-& {\theta(j,l,p,k+1)\, \bK(\bsh_{j,k+1})\over\sqrt{{\vert j-k-1 \vert}^3}}
-{\theta(j+1,l,p,k) \,\bK(\bsh_{j+1,k})\over\sqrt{{\vert j-k+1 \vert}^3}}
\, \Biggr\rbrack
\label{Gam}
\end{eqnarray}
while the function $\bK(\bs)$ is defined by the equation, 
\begin{equation}
\bK(\bs)={-2 \over (2 \pi)^{3/2}} \int d {\bfk} \,{1 \over k^2} \,{\bfk}\,
\biggl( \bu - {{\bfk}{\bfk} \over k^2} \biggr) \,{\bfk} \,\,
\exp( - { 1 \over 2} \, {\bfk} \cdot \bs \cdot {\bfk}) 
\label{kfunc}
\end{equation}
The function $\theta(j,l,p,k)$ is unity if $l$ and $p$ lie
between $j$ and $k$, and zero otherwise,
\begin{equation}
\theta(j,l,p,k)=\cases{1& if $j \leq l,p < k$ \quad or\quad
$k \leq l,p < j$\cr
\noalign{\vskip3pt}
0& otherwise\cr} 
\end{equation}
The convention $\bK(\bsh_{jj})/0=0$, has been adopted in equation~(\ref{Gam}). 
Both the hydrodynamic interaction functions $\bH(\bs)$ and $\bK(\bs)$
can be evaluated analytically in terms of elliptic integrals.
The properties of these functions are discussed in great detail in the
papers by \Ott and coworkers~\cite{ottca2,ottga,ottrab,zylkaga,zylottrg}.

All the approximations discussed earlier (with the exception of 
the decoupling approximation) can be derived by a process of 
succesive simplification of the explicit results for the Gaussian 
approximation given above. The equations that govern the self
consistently averaged theory can be obtained by dropping the last term
in equation~(\ref{gasecmom}), which accounts for the presence of 
fluctuations. Replacing $\bH(\bs)$ by $\bu$ in these truncated equations 
leads to the governing equations of the Zimm model, while setting
$h^* = 0$ leads to the Rouse model.

Material functions predicted by the Gaussian approximation 
in any arbitrary homogeneous flow may be obtained 
by solving the system of  $(N-1)^2$ coupled ordinary differential equations 
for the components of the covariance matrix $\bsjk$~(\ref{gasecmom}). 
Small amplitude oscillatory shear flows and steady shear flow in the
limit of zero shear rate have been examined by \Ott~\cite{ottga} for 
chains with $N \leq 30$ beads, while Zylka~\cite{zylkaga} has
obtained the material functions in steady shear flow for chains 
with $N \leq 15$ beads and compared his results with those
of Brownian dynamics simulations (the comparison was made for chains with
$N=12$ beads). 

The curves predicted by the Gaussian approximation for the storage and 
loss modulus, $G^\prime$ and $G^{\prime\prime}$, as a function of the 
frequency $\omega$, are nearly indistinguishable from the 
predictions of the Zimm theory, suggesting that the Zimm approximation 
is quite adequate for the prediction of linear visco-elastic properties.
There is, however, a significant difference in the prediction of the 
relaxation spectrum. While the Zimm model predicts a set of $(N-1)$ 
relaxation times with equal relaxation weights, the Gaussian approximation 
predicts a much larger set of relaxation times than the number of springs 
in the chain, with relaxation weights that are different and 
dependent on the strength of the hydrodynamic interaction~\cite{ottzylspk}. 
These results indicate that entirely different relaxation spectrum 
lead to similar curves for $G^\prime$ and $G^{\prime\prime}$, and calls into 
question the common practice of obtaining the relaxation spectrum 
from experimentally measured curves for $G^\prime$ and 
$G^{\prime\prime}$ (see also the discussion in~\cite{prakcamb}). 

The zero shear rate viscosity and first normal stress difference 
predicted by the Gaussian approximation are found to be smaller than the
Zimm predictions for all chain lengths. By extrapolating finite chain
length results to the infinite chain limit, \Ott has shown that this 
reduction is by a factor of 72\% -- 73\%, independent of the strength of 
the \hi parameter. Other universal ratios predicted by the Gaussian 
approximation in the limit of zero shear rate are discussed later 
in the chapter.  

A comparison of the predicted shear rate dependence of  
material functions in simple shear flow with the results of Brownian dynamics 
simulations reveals that of all the approximate treatments of 
hydrodynamic interaction introduced so far, the Gaussian approximation 
is the most accurate~\cite{zylkaga}. Indeed, at low shear rates, the 
{\it negative} second normal stress difference predicted by the
Gaussian approximation is in accordance with the simulations results. 

Inspite of the accuracy of the Gaussian approximation, its main drawback 
is its computational intensity, which renders it difficult to 
examine chains with large values of $N$. Apart from the need to 
examine long chains for the reason cited earlier, it also necessary 
to do so in order to obtain the universal predictions of the model. 
A recently introduced approximation which 
enables the evaluation of universal viscometric functions
in shear flow is discussed in the section below. 
Before doing so, however, we first discuss the significant difference  
that a refined treatment of hydrodynamic interaction makes to the prediction 
of translational diffusivity in dilute polymer solutions. 

The correct prediction of the scaling of the diffusion coefficient 
with molecular weight upon introduction of pre-averaged hydrodynamic 
interaction in the Zimm model demonstrates the significant influence that 
\hi has on the translational diffusivity of the macromolecule. While the
pre-averaging assumption appears adequate at equilibrium, it predicts a 
shear rate independent {\it scalar} diffusivity even in the presence of a 
flow field. On the other hand, both the improved treatments of hydrodynamic 
interaction, namely, consistent averaging and the Gaussian
approximation, reveal that the translational diffusivity of a Hookean 
dumbbell in a flowing homogeneous solution is described by an 
anisotropic diffusion tensor which is flow rate 
dependent~\cite{otttdca,otttdca2,otttdga}. 
Indeed, unlike in the Zimm case, the three different approaches
mentioned earlier for calculating the translational diffusivity do not
lead to identical expressions for the diffusion 
tensor~\cite{otttdca,otttdca2}. Insight into the 
origin of the anisotropic and flow rate dependent behavior of the
translational diffusivity is obtained when the link between the 
polymer diffusivity and the shape of the polymer molecule 
in flow~\cite{hoag} is explored~\cite{prakmas1,prakmas2}. It is found that 
the solvent flow field alters the distribution of mass about the centre of 
the dumbbell. As a consequence, the dumbbell experiences an 
average friction that is anisotropic and flow rate dependent. The discussion of 
the influence of improved treatments of hydrodynamic interaction on the 
translational diffusivity has so far been confined to the Hookean dumbbell 
model. This is because the concept of the center of resistance, which is 
very useful for simplifying calculations for bead-spring chains 
in the Zimm case, cannot be employed in these improved 
treatments~\cite{otttdca}. 

\subsection{The twofold normal approximation} 
The twofold normal approximation borrows ideas from the decoupling 
approximation of Magda {\it et al.}~\cite{magda} and Kishbaugh and 
McHugh~\cite{kish} in order to reduce the computational intensity of 
the Gaussian approximation. As in the case of the 
Gaussian approximation, and unlike in the case of the consistent-averaging and 
decoupling approximations where it is neglected, fluctuations in the 
hydrodynamic interaction are included.  
In a sense, the twofold normal approximation is to the Gaussian 
approximation, what the decoupling approximation is to
the consistent-averaging approximation. The computational efficiency of 
the decoupling approximation is due both to the reduction in 
the set of differential equations that 
must be solved in order to obtain the stress tensor, and to the
procedure that is adopted to solve them~\cite{prakott}. These aspects
are also responsible for the computational efficiency of the twofold 
normal approximation. However, the derivation of the reduced set of
equations in the twofold normal approximation is significantly different
from the scheme adopted in the decoupling approximation; it is more
straight forward, and avoids the inconsistencies that are present in the 
decoupling approximation. 

Essentially the twofold normal approximation, (a) assumes that the 
configurational distribution function $\psi$ is Gaussian, (b) uses the 
Rouse or the Zimm orthogonal matrix $\pijk$ to map
$\bQ_j$ to `normal' coordinates $\bQ_j^\prime$, and (c) assumes that the  
covariance matrix $\bsjk \,$ is diagonalised by the same orthogonal matrix, 
{\it ie.} $\sum_{j, \, k} \, \Pi_{jp}\, \bsjk\, \Pi_{kq}=\avel \bQ_p^\prime
 \bQ_q^\prime \aver=\bs_p^\prime \, \delta_{pq}. $
This leads to the following equations for the 
$(N-1)$ variances $\bs_j^\prime$, 
\begin{equation}
{d \over dt} \bs_j^\prime = \bk \cdot \bs_j^\prime
+ \bs_j^\prime \cdot \bk^T 
+{2 k_B T \over \zeta}\,\, \bL_{j} - {H \over \zeta} \, \biggl\lbrack
\bs_j^\prime \cdot \bL_{j} + \bL_{j} \cdot \bs_j^\prime
 \biggr\rbrack 
- {H \over \zeta}\,{H \over k_B T}\, \sk \, \biggl\lbrack
\bs_j^\prime \cdot \bDel_{jk} : \bs_k^\prime + \bs_k^\prime : \bDel_{jk} \cdot
\bs_j^\prime \biggr\rbrack 
\label{tfnsecmom}
\end{equation}
where, $\bL_j \equiv {\widetilde \bL_{jj}}$ are the diagonal tensor
components of the matrix ${\widetilde \bL_{jk}}$,
\begin{equation}
{\widetilde \bL_{jk}}=\sum_{l, \, p} \, \Pi_{lj}\, \bAblp\, \Pi_{pk}
\label{Lam}
\end{equation}
and the matrix $\bDel_{jk}$ is given by,
\begin{equation}
\bDel_{jk}=\sum_{l, \, m, \, n, \, p} \, \Pi_{lj}\,\Pi_{pk}\,
\bG_{lp,mn}\, \Pi_{mj}\,\Pi_{nk}
\label{Del}
\end{equation}
In equations~(\ref{Lam}) and~(\ref{Del}), the tensors $\bAbjk$ and 
$\bG_{lp,mn}$ are given by equations~(\ref{Abar}) and~(\ref{Gam}),
respectively. However, the argument
of the hydrodynamic interaction functions is now given by,
\begin{equation}
\bshmn={1 \over \vert \mu - \nu \vert }\, {H \over k_B T }
\,\, \sum_{j,k \, = \, \min(\mu,\nu)}^{ \max(\mu,\nu)-1} \,\, \sm
\Pi_{jm}\, \Pi_{km}\, \bs_m^\prime
\end{equation}

The decoupling approximation is recovered from the twofold normal 
approximation when the last term in equation~(\ref{tfnsecmom}), 
which accounts for fluctuations in hydrodynamic interaction,
is dropped. Thus the two different routes for finding 
governing equations for the quantities $\bs_j^\prime$ 
lead to the same result. However, Prakash and 
{\" O}ttinger~\cite{prakott} have shown that this is in
some sense a fortuitous result, and indeed the key assumption made in the 
decoupling approximation regarding the diagonalisation of $\bAbjk$ 
is not tenable. The Zimm model in terms of normal modes may be obtained from
equation~(\ref{tfnsecmom}) by dropping the last term, and substituting  
${\widetilde A}_{jk}$ in place of $\bAbjk$. Of course the Zimm orthogonal 
matrix must be used to carry out the diagonalisation in equation~(\ref{Lam}). 
The diagonalised Zimm model reduces to the diagonalised Rouse model 
upon using the Rouse orthogonal matrix and on setting $h^* = 0 $. 

The evolution equations~(\ref{tfnsecmom}) have been solved to obtain the
zero shear rate properties for chains with $ N \le 150$, when the 
Zimm orthogonal matrix is used for the purpose of diagonalisation, 
and for chains with $ N \le 400$, when the Rouse orthogonal matrix is 
used. Viscometric functions at finite shear rates in simple shear flows 
have been obtained for chains with $ N \le 100$~\cite{prakott}. The results 
are very close to those of the Gaussian approximation; this implies that 
they must also lie close to the results of exact
Brownian dynamics simulations. The reasons for the reduction in 
computational intensity of the twofold normal approximation 
are discussed in some detail in~\cite{prakott}. The most important 
consequence of introducing the twofold normal approximation is 
that rheological data accumulated for chains
with as many as 100 beads can be extrapolated to the limit $N \to 
\infty$, and as a result, universal predictions may be obtained.

\begin{figure}[t]
\centerline{ \epsfysize=4.5in
 \epsfbox{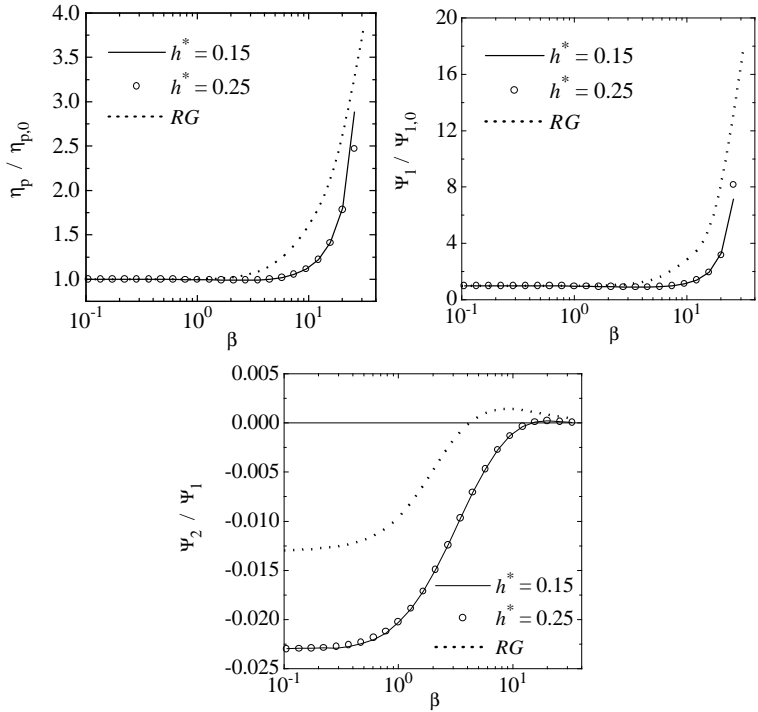} }
\caption{Universal viscometric functions in theta solvents.  
Reproduced from~\protect\cite{prakott}. }
\label{unifig}
\end{figure}

\subsection{Universal properties in theta solvents}
One of the most important goals of examining the influence of hydrodynamic 
interactions on polymer dynamics in dilute solutions is the calculation
of universal ratios and master curves. These properties do not depend 
on the mechanical model used to represent the polymer molecule.   
Consequently, they reflect the most general consequence of the way in which 
\hi has been treated in the theory. They are also the best means to compare 
theoretical predictions with experimental observations since they 
are parameter free. 

There appear to be two routes by which the  
universal predictions of models with hydrodynamic interaction 
have been obtained so far, namely, by extrapolating finite chain 
length results to the limit of infinite 
chain length where the model predictions become parameter free, and
by using renormalisation group theory methods. 

In the former method, there are two essential requirements. 
The first is that rheological data for finite chains must 
be generated for large enough values of $N$ so as to be 
able to extrapolate reliably, {\it ie.} with 
small enough error, to the limit $N \to  \infty$. The second is that 
some knowledge of the leading order corrections to the infinite chain
length limit must be obtained in order to carry out the extrapolation in an 
efficient manner. It is clear from the discussion of the various
approximate treatments of \hi above that it is possible to obtain 
universal ratios in the zero shear rate limit in all the cases. 
Four universal ratios that are 
frequently used to represent the rheological behavior of dilute 
polymer solutions in the limit of zero shear rate are~\cite{ottbook},  
\begin{eqnarray}
U_{\eta \lambda } &=& { \eta_{p, 0} \over {n k_B T \lambda_1} }
\quad \quad \quad \quad \quad \quad 
U_{\eta R}=\lim_{n \to 0} \, {\eta_{p, 0} \over {n \eta_s (4 \pi R_g^3/3)}}
\nonumber \\ 
U_{\Psi \eta } &=& {n k_B T \Psi_{1, 0} \over { \eta_{p, 0}^2}}
\quad \quad \quad \quad \quad \, U_{\Psi \Psi }={\Psi_{2, 0} 
\over \Psi_{1, 0} }
\label{unirat}
\end{eqnarray}
where, $\lambda_1$ is the longest relaxation time, and
$R_g$ is the root-mean-square radius of gyration at equilibrium.
With regard to the leading order corrections to these ratios, it has been  
possible to obtain them explicitly only in the consistently-averaged 
case~\cite{ottca1}. In both the Gaussian approximation and 
the twofold normal approximation it is 
assumed that the leading order corrections are of the same order, 
and extrapolation is carried out numerically by plotting the data as 
a function of $(1/\surd N)$. Because of their computational intensity, 
it is not possible to to obtain the universal shear rate dependence of
the viscometric functions predicted by the consistent-averaging and 
Gaussian approximations. However, it is possible to obtain these 
master curves with the twofold normal approximation. 

Table~\ref{unitab} presents the prediction of the universal 
ratios~(\ref{unirat}) by the various approximate treatments. 
Miyaki {\it et al.}~\cite{miyaki} have experimentally obtained 
a value of $U_{\eta R} = 1.49 \, (6)$ for polystyrene in cyclohexane 
at the theta temperature. Figure~\ref{unifig} displays the viscometric 
functions predicted by the two fold normal approximation. 
The coincidence of the curves for the different values of $h^*$ 
indicate the parameter 
free nature of these results. Divergence of the curves at high 
shear rates implies that the data accumulated for chains 
with $N \le 100$ is insufficient to carry out an accurate 
extrapolation at these shear rates. The incorporation of 
the effect of hydrodynamic interaction into kinetic theory clearly 
leads to the prediction of 
shear thickening at high shear rates even in the long chain limit. 
\begin{table}[bt]
\caption{Universal ratios in the limit of zero shear rate. 
 The exact Zimm values and the Gaussian approximation (GA) values 
for $U_{\eta R}$ and $ U_{\Psi \eta }$ are reproduced 
from~\protect\cite{ottbook}, 
the exact consistent-averaging values 
from~\protect\cite{ottca1}, and 
the renormalisation group (RG) results 
from~\protect\cite{ottrab}. 
The twofold normal approximation values with the Zimm orthogonal 
matrix (TNZ) and the remaining GA values are reproduced 
from~\protect\cite{prakott}. 
Numbers in parentheses indicate the uncertainity in the last figure. }
\label{unitab}
\begin{tabular*}{\textwidth}{@{}l@{\extracolsep{\fill}}llll}
& & & & \\
\hline
&$U_{\eta \lambda}$ & $U_{\eta R}$ & $U_{\Psi \eta}$ & 
$U_{\Psi \Psi}$ \\
\hline
& & & & \\
Zimm & 2.39 & 1.66425 & 0.413865 & 0.0 \\
CA   & 2.39 & 1.66425 & 0.413865 & 0.010628 \\
GA   &$1.835 \, (1)$ & $1.213 \, (3) $ & $0.560 \, (3) $ & 
$ - 0.0226 \, (5)$ \\
RG   & - & 1.377 & 0.6096 & $- 0.0130$ \\
TNZ  &$1.835 \, (1)$ & $1.210 \, (2) $ & $0.5615 \, (3)$ & 
$ - 0.0232 \, (1)$ \\ 
& & & & \\
\hline
\end{tabular*}
\vskip10pt
\end{table}

In both table~\ref{unitab} and figure~\ref{unifig}, the results of 
renormalisation group calculations (RG) are also 
presented~\cite{ottrab,zylottrg}. As mentioned earlier, the 
renormalisation group theory approach is an alternative procedure for 
obtaining universal results. It is essentially a method for refining 
the results of a low-order perturbative treatment of hydrodynamic 
interaction by introducing higher order effects so as to remove the 
ambiguous definition of the bead size. All the infinitely many 
interactions for long chains are brought in through the idea of 
self-similarity. It is a very useful procedure by which a 
low-order perturbation result, which can account 
for only a few interactions, is turned into something meaningful. 
However, systematic results can only be obtained near four dimensions, 
and one cannot estimate the errors in three dimensions reliably. 
The Gaussian and twofold normal approximations on the other hand are 
non-perturbative in nature, and are essentially `uncontrolled' approximations 
with an infinite number of higher order terms. 

It is clear from the figures that the two methods lead to significantly
different results at moderate to high shear rates. A minimum in the 
viscosity and first normal stress difference curves is not predicted
by the renormalisation group calculation, while the twofold normal 
approximation predicts a small decrease from the zero shear rate value
before the monotonic increase at higher shear rates. The good  
comparison with the results of Brownian dynamics simulations for 
short chains indicates that the twofold normal approximation  
is likely to be more accurate than the renormalisation group calculations. 

\section{CONCLUSIONS}
This chapter discusses the development of a unified basis for the treatment 
of non-linear microscopic phenomena in molecular theories of dilute 
polymer solutions and reviews the recent advances in the treatment of 
hydrodynamic interaction. In particular, the successive refinements 
which ultimately lead to the prediction of universal viscometric 
functions in theta solvents have been highlighted.

\end{document}

%% file: defpkt.tex
\def \Ott{ {\"{O}ttinger $\!\!$ }}

\def \bA{ {\widetilde {{\mbox {\boldmath $A$}}} }}
\def \bAb{ {\overline {{\mbox {\boldmath $A$}}} }}
\def \bO{ {{\mbox {\boldmath $\Omega$}}}  }
\def \bu{ {{\mbox {\boldmath $1$}}}   }

\def \bs{ {{\mbox {\boldmath $\sigma \! $}}}}
\def \btau{ {{\mbox {\boldmath $\tau$}}}}
\def \bgam{ {{\mbox {\boldmath $\gamma$}}}}

\def \br{ {{\mbox {\boldmath $r$}}}}
\def \bR{ {{\mbox {\boldmath $R$}}}}
\def \bF{ {{\mbox {\boldmath $F$}}}}
\def \bp{ {{\mbox {\boldmath $p$}}}}
\def \bfu{ {{\mbox {\boldmath $u$}}}}
\def \bzero{ {{\mbox {\boldmath $0$}}}}
\def \bff{ {{\mbox {\boldmath $f$}}}}
\def \bfk{ {{\mbox {\boldmath $k$}}}}

\def \bQ{ {{\mbox {\boldmath $Q$}}}}
\def \bv{ {{\mbox {\boldmath $v$}}}}

\def \bsh{ {\hat \bs}}
\def \bk{ {\mbox {\boldmath $\kappa$}}}
\def \avel { {\bigl\langle}}
\def \aver { {\bigr\rangle}}
\def \bG{ {\bf \Gamma}}
\def \bDel{ {\bf \Delta}}
\def \bL{ {\bf \Lambda}}

\def \lmav{ {
\parallel^{\kern-.40em\raise.08em\hbox{--}}_{\kern-.39em\lower.28em\hbox{--}}}}
\def \rmav{ {
\parallel^{\kern-.59em\raise.08em\hbox{--}}_{\kern-.58em\lower.28em\hbox{--}}}}

\def \bAbjk{ {\bAb_{jk}}}
\def \bAblp{ {\bAb_{lp}}}

\def \bsjk{ {\bs_{jk}}}
\def \bshmn{ {\hat \bs_{\mu \nu}}}
\def \bH{ {{\mbox {\boldmath $H$}}}}
\def \bK{ {{\mbox {\boldmath $K$}}}}
\def \sj{ {\sum_j \, }}
\def \sl{ {\sum_l \, }}

\def \sk{ {\sum_k \,}}
\def \sm{ {\sum_m \,}}

\def \pijk{ {\Pi_{jk}}}
\def \pimj{ {\Pi_{mj}}}
\def \pimk{ {\Pi_{mk}}}

\def \hi{ {hydrodynamic interaction }}